\documentclass[11pt]{amsart}
\usepackage{amsmath, amssymb}
\usepackage{amsfonts}
\usepackage{mathrsfs,mathtools}
\usepackage{graphics,color}
\usepackage{nicefrac}
\newtheorem{theorem}{Theorem}[section]

\newtheorem{proposition}[theorem]{Proposition}
\newtheorem{corollary}[theorem]{Corollary}
\newtheorem{lemma}[theorem]{Lemma}
\newtheorem{example}[theorem]{Example}

\newtheorem{algorithm}[theorem]{Algorithm}

\def\qed{\hfill $\Box$\medskip}
\def\IC{{\mathbb C}}
\def\IR{{\mathbb R}}

\def\cS{{\mathcal S}}

\def\cL{{\mathcal L}}

\def\b11{{\bf 1}}
\def\b00{{\bf 0}}
\def\({\left (}
\def\){\right )}

\def\tr{{\rm tr}\,}
\def\rank{{\rm rank}\,}

\def\diag{{\rm diag}\,}

\def\cL{{\mathcal L}}
\def\Row{{\rm Row}}

\usepackage{amsaddr}

\makeatletter
\renewcommand{\email}[2][]{%
  \ifx\emails\@empty\relax\else{\g@addto@macro\emails{,\space}}\fi%
  \@ifnotempty{#1}{\g@addto@macro\emails{\textrm{(#1)}\space}}%
  \g@addto@macro\emails{#2}%
}
\makeatother

\title{Construction of quantum states with special properties by projection methods}

\author{Xuefeng Duan}
\address[A1]{College of Mathematics and Computational Science,
Guilin University of Electronic Technology, Guilin 541004, P.R. China.\\
\emph{\texttt{duanxuefeng@guet.edu.cn}}}
\author{Chi-Kwong Li}
\address[A2]{Department of Mathematics, College of William and Mary, Williamsburg, Virginia, 23185, USA\\
\emph{\texttt{ckli@math.wm.edu}}}
\author{Diane Christine Pelejo}
\address[A3]{
Institute of Mathematics, College of Science, University of the Philippines Diliman, Diliman, Quezon City 1101, Philippines\\
\emph{\texttt{dcpelejo@math.upd.edu.ph}}}
\begin{document}
\openup .55\jot

\begin{abstract}
We use projection methods to construct (global) quantum states with prescribed reduced (marginal) states, and possibly with some special properties such as having specific eigenvalues, having specific rank and extreme von Neumann or R\'{e}nyi entropy. Using convex analysis, optimization techniques on matrix manifolds, we obtain algorithms to solve the problem. Matlab programs are written based on these algorithms and numerical examples are illustrated. The numerical results reveal new patterns leading to new insights and research problems on the topic.
\end{abstract}

\maketitle



\section{Introduction}

In quantum information science, quantum states are used to store,
process, and transmit information. Mathematically,  quantum states
are represented by density matrices, i.e., positive semidefinite
matrices of trace 1; for example see \cite{K,NC}. Thus, many problems
in quantum information science are connected to the study of density 
matrices and transformations on density matrices with special properties.

Let $M_n$ be the set of $n\times n$ complex matrices, $H_n$ be the set of all $n\times n$ Hermitian matrices and let
$D_n$ be the set of all $n\times n$ density matrices. Consider $k$ quantum systems $\mathcal{X}_1, \ldots, \mathcal{X}_k$ 
with states $\rho_1 \in D_{n_1}, \rho_2\in D_{n_2}, \ldots, \rho_k\in D_{n_k}$, respectively. Their product state is given by
\begin{equation} \label{product}
\rho_1 \otimes \cdots \otimes \rho_k\in D_{n_1 \cdots n_k},
\end{equation}
which is the state of the $k$-partite system $\mathcal{X}=(\mathcal{X}_1,\ldots, \mathcal{X}_k)$ if the $k$ systems are independent. In general, however, the state of $\mathcal{X}$ is a density matrix $\rho \in D_{n_1 \cdots n_k}$, which may not be expressible in the form (\ref{product}). From the state $\rho$ of $\mathcal{X}$, one may extract information about the state of any of its subsystems using a family of linear maps called the \textit{partial trace maps}. To define these maps, note that tensor products like that of equation (\ref{product}) form a spanning set for $H_{n_1 \cdots n_k}$ over the real field and  for $M_{n_1\cdots n_k}$ over the complex field. For a given positive integer $k$, set $\underline{\mathbf{k}}=\{1,\ldots, k\}$ and for any subset $\emptyset \neq J=\{j_1,\ldots,j_r\} \subset \underline{\mathbf{k}}$, let $J^c = \underline{\mathbf{k}} \setminus J$. The \textit{partial trace map with respect to $J$} is the unique linear map $\tr_{J^c}: M_{n_1 \cdots n_k} \longrightarrow M_{n_{j_1}\cdots n_{j_r}}$ such that
\begin{equation}
\tr_{J^c}(\rho_1 \otimes \cdots \otimes \rho_k) =\rho_{j_1}\otimes \cdots \otimes \rho_{j_r}\qquad 
\forall \ \rho_1\otimes \cdots \otimes \rho_k\in D_{n_1,\ldots,n_k} 
\end{equation}
If $\rho$ is the state of the $k-$partite system $\mathcal{X}=(\mathcal{X}_1,\ldots, \mathcal{X}_k)$, then $\tr_{J^c}(\rho):=\rho_J$ is called the reduced state of the subsystem indexed by $J$, i.e. $(\mathcal{X}_{j_1},\ldots, \mathcal{X}_{j_r})$. For completeness, note that if $J=\emptyset$, then we can take $\mbox{tr}_J$ to be the identity map and if $J=\underline{\mathbf{k}}$, we have $tr_J$ is just the usual trace map. 

For example, if $k = 2$, we have a bipartite system.
There are two partial traces of the form
$$\rho_1\otimes \rho_2 \xmapsto{\tr_2} \rho_1 \quad \hbox{ and } \quad
\rho_1 \otimes \rho_2 \xmapsto{\tr_1} \rho_2$$
for any product states $\rho_1 \otimes \rho_2$.
Here we use the notations $\tr_2$ and $\tr_1$ instead of $\tr_{\{2\}}$ and $\tr_{\{1\}}$ for notation simplicity. It is useful to note that if we partition a density matrix $\rho = [\rho_{ij}]_{ i, j \in \underline{\mathbf{n_1}}} \in D_{n_1\cdot n_2}$
such that $\rho_{ij} \in M_{n_2}$, we have
\[\tr_1(\rho) = \sum_{j=1}^{n_1}\rho_{jj} \in M_{n_2}
\quad \hbox{ and } \quad
\tr_2(\rho) = [\tr \rho_{ij}]_{i, j \in \underline{\mathbf{n_1}}} \in M_{n_1}.\]
If $k = 3$, we have a tripartite system, and
there are six partial traces such that
$$\tr_1(\rho_1 \otimes \rho_2\otimes \rho_3) = \rho_2 \otimes \rho_3, \quad
\tr_2(\rho_1 \otimes \rho_2\otimes \rho_3) = \rho_1 \otimes \rho_3, \quad
\tr_3(\rho_1 \otimes \rho_2\otimes \rho_3) = \rho_1 \otimes \rho_2,$$
$$\tr_{12}(\rho_1 \otimes \rho_2\otimes \rho_3) = \rho_3, \qquad
\tr_{23}(\rho_1 \otimes \rho_2\otimes \rho_3) = \rho_1, \qquad
\tr_{13}(\rho_1 \otimes \rho_2\otimes \rho_3) = \rho_2.$$
%The formulas of these partial traces for a general state $\rho\in D_{n_1n_2n_3}$
%are more complicated.

In this paper, we study the following:

\medskip\noindent
{\bf Problem 1.1}
\it Construct  a global state $\rho\in D_{n_1\cdots n_k}$ with certain prescribed reduced (marginal) states $\rho_{J_1}, \dots, \rho_{J_m}$ and with special properties such as having prescribed eigenvalues, prescribed rank, extreme von Neumann entropy, or extreme R\'{e}nyi entropy.

\rm

\medskip
Note that if the mathematical theory shows that the desired global state exists, then one may design experiments to realize the construction.  Otherwise, one has to modify the requirements so that the construction is realizable. (Of course, establishing a physical realization of quantum states is a different challenge altogether.) Also, one may use the mathematical results to predict the properties of the global state if some properties of the reduced states are given or observed. In the extreme case, one may conclude that there are errors in the measurements of reduced states if the mathematical theory suggests that the desired global state cannot exist.

Let us now go back to the mathematical aspect. For a bipartite system, if $\rho_1 \in D_{n_1}$ and $\rho_2 \in D_{n_2}$, then $\rho = \rho_1\otimes \rho_2 \in M_{n_1n_2}$ is a global state having reduced states $\rho_1$ and $\rho_2$. However, it is not easy to construct a global state with prescribed eigenvalues. Researchers have used advanced techniques in representation theory (see \cite{Hy,Kly2} and their references) to study the eigenvalues of the global state and the reduced states. The results are described in terms of numerous linear inequalities even for a moderate size problem (see \cite{Kly2}). Moreover, even if one knows that a global state with prescribed eigenvalues exists, it is not possible to construct the density matrix based on the proof. It is not easy to use these results to answer basic problems, test conjectures, or find general patterns of global states with prescribed properties. For a multipartite system with more than two subsystems, the problem is more challenging. Not much results are available. For example, for a tripartite system, determining whether there is a state $\rho \in D_{n_1n_2n_3}$ with given reduced states $\rho_{12} \in D_{n_1n_2}$ and $\rho_{23} \in D_{n_2n_3}$ is an open problem.

\medskip

In this paper, we will use convex analysis, optimization techniques on matrix manifolds, etc. to obtain algorithms to solve the problem by projection methods. Matlab programs are written based on these algorithms and numerical examples are illustrated. The numerical results reveal new patterns leading to new insights and research problems on the topic. Our paper is organized as follows:

In Sections 3-5, we will focus on the bipartite systems and obtain algorithms based on projection methods to solve Problem 1.1. In Section 6, we extended the results to multipartite systems with more than two subsystems. Proofs can be found in Appendices \ref{proofs1}-\ref{proofs6} and numerical examples are given in Section 7, to illustrate the algorithms used.

\section{Preliminaries}\label{prelim}

Let $\rho_1\in D_{n_1}$ and $\rho_2\in D_{n_2}$.
For bipartite states, we consider the set
\begin{equation}
\cS(\rho_1,\rho_2) = \{\rho\in  D_{n_1\cdot n_2}:
\tr_1(\rho) = \rho_2, \tr_2(\rho) = \rho_1\}.
\end{equation}
Evidently, the set
$\cS(\rho_1,\rho_2)$ is {\it compact, convex, and non-empty}
containing  $\rho_1\otimes \rho_2$. Note that if $T: M_{n_1n_2}\longrightarrow M_{n_1n_2}$ is the linear map satisfying $T(X_1\otimes X_2)=X_2\otimes X_1$ for any $X_1\in M_{n_1}$ and $X_2\in M_{n_2}$, then
\[\cS(\rho_2 , \rho_1)
= \left\{ T(\rho): \rho \in \cS(\rho_1,\rho_2)\right\}.\]
When proving properties of $\mathcal{S}(\rho_1,\rho_2)$, we will often use this fact to assume without loss of generality that $n_1\leq n_2$. Additionally, we can also focus on the case when $\rho_1$ and $\rho_2$ are diagonal density matrices since for any unitary $U \in M_{n_1}$ and $V \in M_{n_2}$, 
\[\cS(U\rho_1U^*, V \rho_2 V^*)
= \{ (U\otimes V)\rho (U\otimes V)^*: \rho \in \cS(\rho_1,\rho_2)\}
= (U\otimes V)\cS(\rho_1,\rho_2) (U\otimes V)^*.\]
Lastly, if $\rho_1$ and $\rho_2$ are nonsingular, then we can translate properties of $S(\rho_1,\rho_2)$ to the general case using the fact that
\[\cS(\rho_1\oplus 0_s , \rho_2 \oplus 0_t)
= \left\{ [\rho_{ij}\oplus 0_t]_{i, j \in \underline{\mathbf{n_1}}}\oplus 0_{s(n_2+t)}: \rho_{ij}\in M_{n_2} \forall i,j \mbox{ and } [\rho_{ij}]_{i, j \in \underline{\mathbf{n_1}}} \in \cS(\rho_1,\rho_2)\right\}.\]

Similarly, for a tripartite system, one may assume that $\mbox{tr}_{12}(\rho)$, $\mbox{tr}_{23}(\rho)$, $\mbox{tr}_{13}(\rho)$ are diagonal matrices, but $\mbox{tr}_{1}(\rho)$, $\mbox{tr}_{2}(\rho)$, $\mbox{tr}_{3}(\rho)$ may be full matrices. So, the study is more intricate. 

We will use the alternating projection methods to do our constructions. The basic set up of the method (See \cite{Boyle-Dykstra}) is to define two closed sets  $\Omega_1,\Omega_2$ of Hermitian matrices such as the set of positive semidefinite matrices and the set of Hermitian matrices having the desired partial traces. Then start with an element $A_0$, say, in  $\Omega_1$. For $m\geq 0$ determine the element $B_m\in \Omega_2$ nearest to $A_m$ and then determine the element $A_{m+1}\in \Omega_1$ nearest to $B_m$. It is known that if $\Omega_1$, $\Omega_2$ are convex (or having other nice properties), then $\lim\limits_{m\rightarrow \infty} A_m = A_*\in \Omega_1$ and $\lim\limits_{m\rightarrow \infty} B_m = B_*\in \Omega_2$ so that $A_*$ and $B_*$ attains the minimum distance between the two sets. In particular, $A_* = B_*\in \Omega_1\cap \Omega_2$ if the two sets have a non-empty intersection.

\section{Bipartite States: Global State with Prescribed Eigenvalues}
In this section, we will consider the problem of finding $\rho\in \mathcal{S}(\rho_1,\rho_2)$ having a prescribed set of eigenvalues $(c_1,\ldots,c_{n_1n_2})$. The problem has been studied by other researchers motivated by problems in quantum chemistry; see for example \cite{Kly2,Fulton,Kly1}, but it is difficult to get a nice theoretical answer. As mentioned in the introduction, by the existing results, even if we know that such a $\rho$ exists, it is difficult to construct the desired density matrix. We will use projection methods to solve the problem as follows.
Let $\rho_1 \in D_{n_1}$ and $\rho_2 \in D_{n_2}$ be density matrices and $c_1\geq \cdots \geq c_{n_1n_2}$. Define the sets $\Omega_1$ and $\Omega_2$ as follows
\begin{equation}\label{omega1}
\Omega_1 = \left\{\rho = [\rho_{ij}]\in M_{n_1}(M_{n_2}):
\sum\limits_{i=1}^{n_1} \rho_{ii} = \rho_2,\ [\tr \rho_{ij}]_{i, j \in \underline{\mathbf{n_1}}} = \rho_1\right\}
\end{equation}
\begin{equation}\label{omega2}
\Omega_2 = \{W \diag(c_1, \dots, c_{n_1n_2})W^*: W\in M_{n_1n_2} \hbox{ is unitary}\}.
\end{equation}
We consider the two projection
operators $\Phi_{\Omega_1}: H_{n_1n_2} \longrightarrow \Omega_1$ and $\Phi_{\Omega_2}: H_{n_1n_2} \longrightarrow \Omega_2$. That is,
\[||P-\Phi_{\Omega_1}(P)||=\min\limits_{Z\in \Omega_1} ||P-Z||\quad \mbox{ and }\quad ||P-\Phi_{\Omega_2}(P)||=\min\limits_{Z\in \Omega_2} ||P-Z|| \]

We can determine $\Phi_{\Omega_2}$ using the following result; for example, see \cite[Theorem 10.B.10]{maj}.

\begin{theorem}\label{2.1}
Let $\|\cdot\|$ be a unitary similarity invariant
norm,  i.e., $\|X\| = \|W^*XW\|$ for any $X \in H_N$ and unitary $W \in M_N$. 
 Suppose $P = UDU^* \in H_{N}$,
where $U\in M_{N}$ is unitary and $D$ is a diagonal matrix
with diagonal entries arranged in descending order. Then,
$$\|P - U\diag(c_1, \dots, c_{n_1n_2})U^*\| \le \|P-Z\| \qquad \hbox{ for all } Z \in \Omega_2.$$ 
\end{theorem}

In our study, we always use the Frobenius norm $\|X\| = [\tr(X^*X)]^{1/2}$, which is unitary similarity invariant. By the above theorem, we have 
\begin{equation}\label{proj2e}
\Phi_{\Omega_2}(P) = U \diag(c_1, \dots, c_{n_1n_2})U^* \quad \hbox{ if }
P = U\diag(\mu_1, \dots, \mu_{n_1n_2})U^* \hbox{ with } \mu_1 \ge \cdots \ge \mu_{n_1n_2}.
\end{equation}
Here $U$ may not be unique if $P$ has repeated eigenvalues, in which case, we just choose any one of them.

The next proposition, whose proof can be seen in Appendix \ref{proofs1}, provides an explicit formula for $\Phi_{\Omega_1}(P)$ using the Karush$-$Kuhn$-$Tucker (KKT) conditions. It connects our problem to other optimization problems. The result will also follow from Proposition \ref{multiprojform}, which covers the more general multipartite systems. 

\begin{proposition}\label{proj1}  Given a block matrix $P=[P_{ij}]\in M_{n_1}(M_{n_2})$,
the projection operator of $P$ onto $\Omega_{1}$ is given by
\begin{equation}\label{proj1e}
\Phi_{\Omega_1}(P)=P-\frac{I_{n_1}}{n_1}\otimes \Big(\mbox{\rm tr}_1(P)-\rho_2\Big)-\Big(\mbox{\rm tr}_2(P)-\rho_1\Big)\otimes \frac{I_{n_2}}{n_2}+\frac{\mbox{\rm tr}(P)-1}{n_1n_2}I_{n_1n_2}
\end{equation}
%\begin{equation}\label{proj1e}
%\Phi_{\Omega_1}(P)=(X_{ij}),\ \ X_{ij}=P_{ij}+\delta_{ij}\frac{1}{n_1}\left(\rho_2-\sum\limits_{i=1}^{n_1}P_{ii}\right)+\frac{1}{n_2}\left(a_{ij}-\mbox{tr}(P_{ij})+\delta_{ij}\frac{tr(P)-1}{n_1}\right)I_{n_2}
%\end{equation}
%where $a_{ij}$ is the $(i,j)^{th}$ entry of $\rho_1$.
\end{proposition}

Using equations (\ref{proj2e}) and (\ref{proj1e}), we can implement the following alternating projection algorithm to find $\rho\in \cS(\rho_1,\rho_2)$ with prescribed eigenvalues $(c_1,\ldots,c_{n_1n_2})$, if it exists.\bigskip

\noindent
\fbox{
\begin{minipage}{0.98\textwidth}
\begin{algorithm}{\textsc{Alternating projection scheme to find 
$\rho=\Phi_{\Omega_1\cap\Omega_2}(X_0)$}}\label{2.3}\bigskip\\
\begin{tabular}{ll}
{\bf Step 1.} &  Generate a random unitary $U$ and a random probability vector 
$(d_1,\ldots,d_{n_1\ldots n_k})$\medskip\\
& and set the initial point to be 
$X_0=U\diag(d_1,\ldots, d_{n_1\ldots n_k})U^*$.\medskip\\ & Choose an integer $N$ (iteration limit) 
and a small positive number $\delta$ (tolerance).\bigskip\\
{\bf Step 2.} & For $k=1,\ldots,N$, define 
$X_{2k-1}=\Phi_{\Omega_1}(X_{2k-2})\mbox{ and } X_{2k}=\Phi_{\Omega_2}(X_{2k-1})$\medskip\\
& If $||{\rm \tr}_1(X_{2k})-\rho_2||+||{\rm \tr}_2(X_{2k})-\rho_1||<\delta$, then declare $X_{2k}$ to be a solution.
\end{tabular}
\end{algorithm}
\end{minipage}}
\ \\

If $\Omega_1\cap \Omega_2\neq \emptyset$, Theorem 4.3 of \cite{Lewis2} guarantees local convergence of this algorithm. That is, if we choose a suitable starting point $X_0$, then the algorithm produces a sequence $\{X_k\}$ that converges to a $\rho\in \Omega_{1}\cap \Omega_{2}$ as $k\longrightarrow \infty$. If $\Omega_1 \cap \Omega_2 = \emptyset$, then $\{X_{2k}\}$
converges to a global state nearest to a matrix with the desired eigenvalues $c_1 \ge \cdots \ge c_{n_1 n_2}$. In case one needs the desired global state to perform some quantum task, then one needs to adjust specifications of the eigenvalues of the global state or the reduced states.

\section{Bipartite States: Global States with Prescribed Ranks}\label{2.2} In this section, we discuss methods to find a low rank solution $\rho\in \cS(\rho_1,\rho_2)$. Such low rank solutions are of great interest as they are often entangled \cite[Theorem 8]{ruskai}. In fact, if $\rank(\rho)$ is strictly less than $\max\{\rank(\rho_1),\rank(\rho_2)\}$, it was shown in \cite[Theorem 1]{horodecki} that $\rho$ must be distillable. It is also known (for example, see \cite{watrous}) that if $\rho \in \mathcal{S}(\rho_1,\rho_2)$, then 
\[ \max\left\{\left\lceil\frac{\rank(\rho_2)}{\rank(\rho_1)}\right\rceil,\left\lceil\frac{\rank(\rho_1)}{\rank(\rho_2)}\right\rceil\right\}\leq \rank(\rho)\leq \rank(\rho_1)\rank(\rho_2)\]
The upper bound is always attained by $\rho=\rho_1\otimes \rho_2$ but the lower bound is not always attained. For example, in \cite[Subsection 3.3.1]{Kly1}, it was shown that there exists a rank one $\rho\in \mathcal{S}(\rho_1,\rho_2)$ if and only if $\rho_1$ and $\rho_2$ are isospectral, that is, $\rho_1$ and $\rho_2$ have the same set of nonzero eigenvalues, counting multiplicities.

The following algorithm is an implementation of an alternating projection method to find a low rank solution $\rho\in \mathcal{S}(\rho_1,\rho_2)$, if it exists. Convergence of this algorithm is not guaranteed but numerical results shown in Section 4 illustrate that this algorithm is effective in finding a low rank solution. \bigskip

\noindent\fbox{
\begin{minipage}{0.98\textwidth}
\begin{algorithm}{\sc Alternating projection scheme to find $\rho\in \mathcal{S}(\rho_1,\rho_2)$ with rank$(\rho)\leq r$.}\label{altprojrank}\bigskip\\
\begin{tabular}{ll}
\textbf{\emph{Step 1:}} 
& Set $k=0$ and choose $X_0\in D_{n_1n_2}$ and a positive integer $N$ (iteration limit) and \\
& a small positive integer $\delta$ (tolerance). Do the next step for $k=1,\ldots, N$.\medskip\\
\textbf{\emph{Step 2:}} & Define $\rho^{(2k-1)}=\Phi_{\Omega_1}(\rho^{(2k-2)})$. If $\rho^{(2k-1)}=U\diag(d_1,\ldots,d_{n_1n_2})U^*$ for some \medskip\\
& unitary $U$ and $d_1\geq d_2 \geq \cdots \geq d_{n_1n_2}\geq 0$, define 
$\rho^{(2k)}=U(s_1,\ldots, s_{r},0,\ldots,0)U^*,$\medskip\\
& where $s_i=\max\{d_i,0\}$.\medskip\\
&  If $||{\rm \tr}_1(X_{2k})-\rho_2||+|{\rm \tr}_2(X_{2k})-\rho_1||< \delta$, then declare $\rho^{(2k)}$ as a solution.
\end{tabular}
\end{algorithm}
\end{minipage}}
\ \\

In view of the fact that the above algorithm may not converge and multiple low rank solutions may exist, we derive other methods to find low rank solutions. Additionally, as we will see in Section \ref{numex}, two of the algorithms produce a solution with low von Neumann entropy.

First, we present the following theorem found in \cite{Kly1} to construct a rank one solution $\rho\in \mathcal{S}(A,B)$ for isospectral $A$ and $B$. Based on this, we present three methods to find a low rank  solution $\rho\in \mathcal{S}(\rho_1,\rho_2)$. 
 
\begin{theorem}\label{pure} Let $\rho_1\in D_{n_1}$ and $\rho_2\in D_{n_2}$ have spectral decomposition $\rho_1=\gamma_1 x_1x_1^*+\cdots + \gamma_k x_kx_k^* $ and $\rho_2=\gamma_1 y_1y_1^*+\cdots + \gamma_k y_ky_k^*$, and 
\[w=\sum\limits_{i=1}^k \sqrt{\gamma_i}(x_i\otimes y_i)\]
Then $P=ww^*\in \mathcal{S}(\rho_1,\rho_2)$. 
\end{theorem}

In the first algorithm that we will present, we can choose an integer $k$ with
\[\max\{\rank(\rho_1),\rank(\rho_2)\} \leq k\leq \rank(\rho_1)+\rank(\rho_2)-1\] and construct a $\rho \in \cS(\rho_1,\rho_2)$ with $\rank(\rho)=k$. We do this by expressing both $\rho_1$ and $\rho_2$ as an average of $k$ pure states (see proof of Proposition \ref{rankk} in Appendix \ref{proofs2}).  \bigskip

\noindent\fbox{
\begin{minipage}{0.98\textwidth}
\begin{algorithm}\label{algrankk}{\sc Construction of a rank $k$ state $\rho\in \cS(\rho_1,\rho_2)$ for any $k$ such that $\max\{\rank(\rho_1),\rank(\rho_2)\}\leq k \leq \rank(\rho_1)+\rank(\rho_2)-1$}\bigskip\\
\begin{tabular}{ll}
\textbf{\emph{Step 1:}} &  Find unitaries $U$ and $V$ such that $\rho_1=U\diag(a_1,\ldots, a_{n_1})U^*$ and \\
& $\rho_2=V\diag(b_1,\ldots,b_{n_2})V^*$.\medskip\\ 
\textbf{\emph{Step 2:}} & Choose an integer $k$ with $\max\{\rank(\rho_1),\rank(\rho_2)\} \leq k\leq \rank(\rho_1)+\rank(\rho_2)-1$\\
& and let $\omega_k$ be a principal $k^{th}$ root of unity. For any $i=1,\ldots, k$, define $x_i\in \mathbb{C}^m$\\
& and $y_i\in \mathbb{C}^n$ such that $x_i=[\omega_k^{(j-1)i} \sqrt{a_j}]$ and  $y_i=[\omega_k^{(j-1)i} \sqrt{b_j}]$.\medskip\\
\textbf{\emph{Step 3:}} & Define $\rho=z_1z_1*+ \cdots +z_kz_k^*$ where $z_i=\frac{1}{\sqrt{k}}(Ux_i\otimes Vy_i)$.
\end{tabular}
\end{algorithm}
\end{minipage}}

\begin{proposition}\label{rankk}
For any integer $k$ with $\max\{\rank(\rho_1),\rank(\rho_2)\}\leq k\leq  \rank(\rho_1)+\rank(\rho_2)-1$, Algorithm \ref{algrankk} produces a $\rho\in \mathcal{S}(\rho_1,\rho_2)$ with $\rank(\rho)=k$.
\end{proposition}

In \cite{LPW}, it was proven that if there is a $\rho\in\cS(\rho_1,\rho_2)$ with rank $k$, then there is $\rho\in \cS(\rho_1,\rho_2)$ with $k\leq \rank(\rho)\leq \rank(\rho_1)\rank(\rho_2)$.  The following theorem is a consequence of this but we will give a constructive proof (see Appendix \ref{proofs3}) by induction and using Proposition \ref{rankk}.

\begin{theorem}\label{inductrank}
For any integer $k$ such that $\max\{\rank(\rho_1),\rank(\rho_2)\}\leq k\leq \rank(\rho_1)\rank(\rho_2)$, there exists $\rho\in \cS(\rho_1,\rho_2)$ with $\rank(\rho)= k$.% Moreover, there is a choice for $\rho_1$ and $\rho_2$ such that no $\rho\in \mathcal{D}_{mn}$ with rank$(\rho)<\max\{r,s\}$ satisfies $tr_1(\rho)=\rho_2$ and $tr_2(\rho)=\rho_1$.
\end{theorem}

Note that if $\min\{\rank(\rho_1),\rank(\rho_2)\}=1$, then $\mathcal{S}(\rho_1,\rho_2)=\{\rho_1\otimes \rho_2\}$. Now, what remains to be seen is the case when $\rank(\rho_1),\rank(\rho_2)\geq 2$ and
 \[\max\left\{\left\lceil\frac{\rank(\rho_2)}{\rank(\rho_1)}\right\rceil,\left\lceil\frac{\rank(\rho_1)}{\rank(\rho_2)}\right\rceil\right\}\quad \leq\quad  k\quad  \leq\quad \max\{\rank(\rho_1),\rank(\rho_2)\}-1.\] 
Can we find $\rho\in \mathcal{S}(\rho_1,\rho_2)$ with rank $k$?
 In the next algorithm, we present one more scheme to find a low rank solution $\rho\in \mathcal{S}(\rho_1,\rho_2)$ using the following known result in \cite{golub}.
\begin{theorem}\label{interlace}
Suppose $a_1\geq b_1\geq a_2\geq b_2 \geq \cdots \geq a_n\geq b_n\geq 0$. Define $d=[d_i]\in \mathbb{R}^n$ such that
\[d_i=\left\{\begin{array}{ll}
0 & \mbox{ if $a_i=0$ or $a_j=a_i$ for some $j\neq i$ }\\
\sqrt{\frac{\prod\limits_{j=1}^n (b_j-a_i) }{-\prod\limits_{\stackrel{j=1}{j\neq i}}^n (a_j-a_i)}} & \mbox{ otherwise}
\end{array}  \right.  \]
Then $\diag(a_1,\ldots,a_n)-dd^*$ has eigenvalues $b_1,\ldots, b_n$. 
\end{theorem}
In the following two algorithms, we denote the vector in $\mathbb{C}^{n}$ having its $j^{th}$ entry equal to one and all other entries equal to zero by $e_j^{(n)}$. 

\noindent\fbox{
\begin{minipage}{0.98\textwidth}
\begin{algorithm}{\sc Construction of $\rho\in \cS(\rho_1,\rho_2)$ with $\rank(\rho)\leq\max\{\rank(\rho_1),\rank(\rho_2)\}$.} \label{alginterlace}\ \bigskip\\
\begin{tabular}{ll}
{\rm \textbf{Step 1:}} & Set $A_0=\rho_1$ and $B_0=\rho_2$. Do the next step with the initial value of $r$ set to $0$. \medskip\\
{\rm \textbf{Step 2:}} & If $A_r=0$, then proceed to step 3, setting $k$ to be equal to the terminal value of $r$.\\
& Otherwise do the following steps. \medskip\\
{\rm \textbf{Step 2.1:}} & Find unitary $U,V$ such that 
$A_r=U(S_1\oplus \cdots \oplus S_p\oplus T_1\oplus T_q\oplus L_a)U^*$ and \\
& 
$B_r=V(\tilde{S}_1\oplus \cdots \oplus \tilde{S}_p\oplus \tilde{T}_1\oplus \tilde{T}_q\oplus L_b)V^*$, where \medskip\\
& \emph{(1)} $T_j=\diag(c_{j1},\ldots,c_{jt_j})$ and $\tilde{T}_j=\diag(d_{j1},\ldots,d_{jt_j})$ \\
& satisfy $d_{j1}\geq c_{j1}\geq \cdots \geq d_{jt_j}\geq c_{jt_j}$,\medskip\\
& \emph{(2)} $S_i=\diag(a_{i1},\ldots,a_{is_i})$ and $\tilde{S}_i=\diag(b_{i1},\ldots,b_{is_i})$\\
& satisfy $a_{i1}\geq b_{i1}\geq \cdots \geq a_{is_i}\geq b_{is_i}$,\medskip\\
& and either $L_a$ is empty or is a zero block or $L_b$ is empty or is a zero block.\medskip\\
{\rm \textbf{Step 2.2:}} &  For $i=1,\ldots, p$, use Theorem \ref{interlace} to find $x_i\in \mathbb{R}^{s_i}$ such that the eigenvalues of\\
& $S_i-x_ix_i^*$ are the eigenvalues of  $\tilde{S}_i$.\\
& Similarly, for $j=1,\ldots, q$, find $y_j\in \mathbb{R}^{t_j}$ such that the eigenvalues of $\tilde{T}_j-y_iy_i^*$\\
&  are the same as that of $T_j$. \medskip\\
{\rm \textbf{Step 2.3:}} &  Let  $C_{r+1}=U\Big(
(S_1 -x_1x_1^*)\oplus \cdots \oplus (S_p-x_px_p^*)\oplus T_1 \oplus \cdots \oplus T_q \oplus 0\Big)U^*$ \\
& and $\tilde{C}_{r+1}=V\Big(
\tilde{S}_1\oplus \cdots \oplus \tilde{S}_p\oplus (\tilde{T}_1-y_1y_1^*) \oplus \cdots \oplus (\tilde{T}_q-y_qy_q^*) \oplus 0\Big)V^*$ \\
& and set  $A_{r+1}=A_r-C_{r+1}$ and $B_{r+1}=B_r-\tilde{C}_{r+1}$. \\
& Increment the value of $r\leftarrow r+1$ and repeat step 2. \medskip\\
{\rm \textbf{Step 3:}} & For $i=1,\ldots,k$, find $U_i$ and $V_i$ such that $C_i=U_i\diag(\alpha_{i1},\ldots,\alpha_{ir_i},0,\ldots)U_i^*$\\
&  and $\tilde{C}_i=V_i\diag(\alpha_{i1},\ldots,\alpha_{ir_i}, 0,\ldots)V_i^*$. Define $w_i=\sum\limits_{j=1}^{r_i} \sqrt{\alpha_{ij}} (U_ie_{j}^{(n_1)}\otimes V_ie_{j}^{(n_2)})$.\medskip\\ 
{\rm \textbf{Step 4:}} & Define $\rho=w_1w_1^*+\cdots + w_kw_k^*$.
\end{tabular} 
\end{algorithm}
\end{minipage}}\medskip

The proof of the following proposition can be found in Appendix \ref{proofs4}.
\begin{proposition}\label{interlacethm}
Let $\rho_1=U\diag(a_1,\ldots, a_{n_1})U^*$ and $\rho_2=V\diag(b_1,\ldots, b_{n_1})V^*$. Algorithm \ref{alginterlace} produces positive semidefinite matrices $C_1,\ldots, C_k\in M_{n_1}$ and $\tilde{C}_1,\ldots, \tilde{C}_k\in M_{n_2}$ such that
\begin{enumerate}
\item  $k\leq \max\{\rank(\rho_1),\rank(\rho_2)\}$,
\item For $i=1,\ldots, k$, the matrices $C_i$ and $\tilde{C}_i$ are isospectral,
\item $\rho_1=C_1+\cdots+C_k$ and $\rho_2=\tilde{C}_1+\cdots + \tilde{C}_k$, 
\item $\tr_1(w_iw_i^*)=\tilde{C}_i$ and $\tr_2(w_iw_i^*)=C_i$ for $i=1,\ldots k$ so that $\rho\in \mathcal{S}(\rho_1,\rho_2)$ and;
\item If $a_{i_1}\geq b_{j_1}\geq \cdots \geq a_{i_l}\geq b_{j_l}$ (or $b_{j_1}\geq a_{i_1}\geq \cdots \geq b_{j_l}\geq a_{i_l}$)  for some distinct indices $i_1,\ldots,i_{l+1}$ and distinct $j_1,\ldots,j_{l+1}$, then the solution $\rho$ produced by Algorithm \ref{alginterlace} has rank at most $\max\{\rank(\rho_1)-l+1,\rank(\rho_2)-l+1\}$. 
\end{enumerate}
\end{proposition}

Finally, we present one more scheme to find a low rank solution $\rho\in\mathcal{S}(\rho_1,\rho_2)$. Similar to Algorithm \ref{alginterlace}, we find $\rho$ by first writing \[\rho_1=C_1+\cdots+ C_k \quad \mbox{ and } \quad \rho_2=\tilde{C}_1+\ldots+\tilde{C}_k\]
for $k$ pairs $(C_1,\tilde{C}_1),\ldots,(C_k,\tilde{C}_k)\in M_{n_1}\times M_{n_2}$, of isospectral positive semidefinite matrices with $k\leq \max\{\rank(\rho_1),\rank(\rho_2)\}$. In fact, these pairs can be chosen so that we can construct a $\rho\in \mathcal{S}(\rho_1,\rho_2)$ whose nonzero eigenvalues are given by $\lambda_i=\mbox{tr}(C_i)=\mbox{tr}(\tilde{C}_i)$ for $i=1,\ldots, k$. Furthermore, this solution $\rho$  satisfies \[||\rho||_{2}=\max\limits_{\sigma\in\mathcal{S}(\rho_1,\rho_2)} ||\sigma||_2,\] where $||\cdot ||_2$ denotes the operator/spectral norm.\bigskip

\noindent\fbox{
\begin{minipage}{0.98\textwidth}
\begin{algorithm}\label{lowrankC} {\sc Construction of $\rho\in \cS(\rho_1,\rho_2)$ with $\rank(\rho)\leq\max\{\rank(\rho_1),\rank(\rho_2)\}$.}\bigskip \\
\begin{tabular}{ll}
\textbf{\emph{Step 1:}} & Suppose $\rho_1=U\diag(a_1,\ldots,a_{n_1})U^*$ and $\rho_2=V\diag(b_1,\ldots,b_{n_2})V^*$. \\
& Set $a_i^{(0)}=a_i\ \forall i\in \underline{\mathbf{n}_1}$  and $b_j^{(0)}=b_j\ \forall j\in \underline{\mathbf{n}_2}$.\\
&  Do the next step with the initial value of $r$ set to $0$.  \medskip\\
\textbf{\emph{Step 2:}} & If $\sum\limits_{i=1}^{n_1} a_i^{(r)}=0$, then proceed to step 3, setting $k$ to be equal to the terminal value of $r$.\\
& Otherwise, find permutation maps $\sigma_r: \underline{\mathbf{n_1}}\rightarrow \underline{\mathbf{n_1}}$ and $\tilde{\sigma}_r:\underline{\mathbf{n_2}} \rightarrow \underline{\mathbf{n_2}}$ such that \medskip\\  
&  $a_{\sigma_r(1)}^{(r)}\geq \cdots \geq a_{\sigma_r(n_1)}^{(r)}$  and $b_{\tilde{\sigma}_r(1)}^{(r)}\geq \cdots \geq b_{\tilde{\sigma}_r(n_2)}^{(r)}$. \medskip\\
& Denote by $P_r$ and $\tilde{P}_r$ the  permutation matrices satisfying \medskip\\
& $P_r\diag(a_1^{(r)},\ldots,a_{n_1}^{(r)})P_r^{T}=\diag(a_{\sigma_r(1)}^{(r)},\ldots,a_{\sigma_r(n_1)}^{(r)})$ and\medskip\\
 & $\tilde{P}_r\diag(b_1^{(r)},\ldots,b_{n_2}^{(r)})\tilde{P}_r^{T}=\diag(b_{\tilde{\sigma}_r(1)}^{(r)},\ldots,b_{\tilde{\sigma}_r(n_2)}^{(k)}). $  \medskip\\
& Define $c^{(r)}_{j}=\min\{a_{\sigma_r(j)}^{(r)},b_{\tilde{\sigma}_r(j)}^{(r)}\}$ if $j\in\{1,\ldots, \min\{n_1,n_2\}\}$ and $c^{(r)}_{j}=0$ otherwise.\medskip\\
& Let $C_r=UP_r^{T}\diag(c^{(r)}_{1},\ldots, c^{(r)}_{n_1})P_rU^*$ and $\tilde{C}_r=V\tilde{P}_r^{T}\diag(c^{(r)}_{1},\ldots, c^{(r)}_{n_2})\tilde{P}_rV^*$. \medskip \\
& Then set $a_i^{(r+1)}=a_i^{(r)}-c^{(r)}_{\sigma_r^{-1}(i)}$ and $b_i^{(r+1)}=b_i^{(r)}-c^{(r)}_{\tilde{\sigma}_r^{-1}(i)}$.\medskip\\
& Increment the value of $r\leftarrow r+1$ and repeat step 2. \medskip\\
\textbf{\emph{Step 3:}} & For $r=1,\ldots,k$, define $w_r=\sum\limits_{j=1}^{\min\{n_1,n_2\}} \sqrt{c^{(r)}_{j}} (Ue_{\sigma_i(j)}^{(n_1)}\otimes Ve_{\tilde{\sigma}_i(j)}^{(n_2)})$\\
&  and $\rho=w_1w_1^*+\cdots + w_kw_k^*$.
\end{tabular}

\end{algorithm}
\end{minipage}}\ \\

The proof of the following proposition can be found in Appendix \ref{proofs5}.
\begin{proposition}\label{lowrankCthm}
Let $\rho_1\in D_{n_1}$ and $\rho_2\in D_{n_2}$. Algorithm \ref{lowrankC} produces positive semidefinite matrices $C_1,\ldots, C_k\in M_{n_1}$ and $\tilde{C}_1,\ldots, \tilde{C}_k\in M_{n_2}$ such that 
\begin{enumerate}
\item $k \leq \max\{\rank(\rho_1),\rank(\rho_2)\}$.
\item For $i=1,\ldots, k$, the matrices $C_i$ and $\tilde{C}_i$ are isospectral.
\item $\rho_1=C_1+\cdots+C_k$ and $\rho_2=\tilde{C}_1+\cdots + \tilde{C}_k$.
\item If $w_1,\ldots, w_k\in\mathbb{C}^{n_1n_2}$ are the vectors defined in Step 3, then $\tr_1(ww_i^*)=\tilde{C}_i$ and $\tr_2(ww_i^*)=C_i$ so that $\rho\in S(\rho_1,\rho_2)$. Moreover,  $w_i^*w_j=\delta_{ij}\tr(C_i)$ so that the nonzero eigenvalues of $\rho$ are precisely $tr(C_1), \ldots, tr(C_k)$.
\item $||\rho||_{2}=\tr(C_1)= \max\limits_{\sigma\in\mathcal{S}(\rho_1,\rho_2)} ||\sigma||_2$.
\end{enumerate}
\end{proposition}

Algorithm \ref{lowrankC} can produce a solution $\rho$ that has rank less than $\min\{\rank(\rho_1),\rank(\rho_2)\}$, but usually does not give the minimum rank. Take for example the case 
\[\rho_1=\diag\left(\frac{7}{10},\frac{3}{10}\right) \mbox{ and } \rho_2=\diag\left(\frac{3}{5},\frac{1}{5},\frac{1}{5}\right).\]
There is no $\rho\in \cS(\rho_1,\rho_2)$ with rank $1$, but there is a rank $2$ solution given by $\rho=w_1w_1^*+w_2w_2^*$, where
\[w_1=\sqrt{\frac{3}{5}} (e_1\otimes e_1)+\sqrt{\frac{1}{10} }(e_2\otimes e_2) \quad \mbox{ and } \quad w_2=\sqrt{\frac{1}{10}} (e_1\otimes e_2)+\sqrt{\frac{1}{5} }(e_2\otimes e_3) \]
However, Algorithm \ref{lowrankC} will produce a rank 3 solution.

The fact that Algorithm \ref{lowrankC} will produce a $C_1$ satisfying Proposition \ref{lowrankCthm}.(5) follows from \cite{Kly1} using algebraic combinatorics. We will give a simple matrix theoretic proof in Appendix \ref{proofs1}. 

Note that the solutions obtained from Algorithms \ref{algrankk}, \ref{alginterlace}, \ref{lowrankC} can be utilized as the starting point when implementing Algorithm \ref{altprojrank} to find a solution with lower rank. As mentioned in the beginning of Subsection \ref{2.2}, finding low rank $\rho\in \mathcal{S}(\rho_1,\rho_2)$ is of interest in the study of distillation. Here, we note that the solution obtained in Algorithm \ref{lowrankC} has relatively low von Neumann entropy since it has maximal spectral norm, that is, its largest eigenvalue is as close to $1$ as possible making it a good pure state approximation. However, as will be seen in the numerical results in Section 3, it is not guaranteed to have minimal von Neumann entropy.  

\section{Bipartite States: Global State with Extremal Entropies}

In this section, we are interested in finding $\rho \in \cS(\rho_1, \rho_2)$ attaining
certain extreme functional values for a given scalar function $f$ on quantum states.
Our result will cover the case when $f(\rho)$ is  the von-Neumann entropy of $\rho$
defined by
\begin{equation}
S(\rho) = -\tr (\rho\log \rho) = -\sum \lambda_j \log(\lambda_j),
\end{equation}
where $\lambda_j$ are the eigenvalues of $\rho$, and $x\log x = 0$ if $x = 0$; and
the R\'{e}nyi entropy defined by
\begin{equation}
 S_\alpha(\rho) =\frac{1}{1-\alpha}\log\tr(\rho^\alpha) = \frac{1}{1-\alpha}
\log\left(\sum \lambda_j^{\alpha} \right)\qquad
\hbox{ for }  \alpha \ge 0.
\end{equation}

Note that $\rho_1\otimes \rho_2 \in \cS(\rho_1,\rho_2)$ has maximum
von Neumann entropy by the subadditivity property of von Neumann entropy.
So, we focus on searching for $\rho \in \cS(\rho_1,\rho_2)$ with
minimum entropy, that is, we are interested in the following minimization problem
\begin{equation}\label{minent}
\min\limits_{\rho\in \Omega_{1}\cap \Omega_{3}} -\tr (\rho\log \rho),
\end{equation} where
\begin{equation}\label{omega3}
\Omega_{3}=\{\rho\in M_{n_1}(M_{n_2}): \rho\geq 0\}.
\end{equation} 
Here $\rho\geq 0$ means that the matrix $\rho$
is positive semidefinite. Since $\Omega_{1}$ and $\Omega_{3}$ are closed convex sets,
then the set $\Omega_{1}\cap \Omega_{3}$ is also a closed convex set. Now we use
the nonmonotone spectral projected gradient ({\bf NSPG}) method to solve
the minimization problem
(\ref{minent}), which was proposed in Birgin et al \cite{Birgin2},
on minimizing a continuously  differentiable function
$f: \IR^{n}\rightarrow \IR$ on a nonempty closed convex set $M.$
As it is quite simple to implement
and very effective for large-scale problems, it has been extensively studied in the past years
(see \cite{Lewis,Lewis2} and their references for details). The NSPG method has the form
$x_{k+1}=x_{k}+\alpha_{k}d_{k},$ where $d_{k}$ is chosen to be
$P_{M}(x_{k}-t_{k}\nabla f(x_{k}))-x_{k}$
with $t_{k}>0$ a precomputed scalar. The direction $d_{k}$ is guaranteed to be a
descent direction (\cite[Lemma 2.1]{Birgin1})
and the step length $\alpha_{k}$ is selected by a nonmonotone linear search strategy. The
key problems when using NSPG method to solve (\ref{minent}) are (i) how to compute the gradient of the objective function
$f(\rho)=-\tr \rho\log \rho$ and (ii) how to determine the projection operator $\Phi_{\Omega_{1}\cap \Omega_{3}}(Z)$ of
$Z$ onto the
set $\Omega_{1}\cap \Omega_{3}.$ Such problems is addressed in the following.

For any function $f: \IR\rightarrow \IR$, one can extend it to
$f: H_n \rightarrow H_n$ such that
$f(A) = \sum f(a_j) P_j$
if $A$ has spectral decomposition  $A = \sum a_j P_j$. where $P_j$ is the
orthogonal projection of $\IC^n$ onto the kernel of $A-a_jI$.
Furthermore, we can consider the scalar function $A \mapsto \tr f(A)$.
By Theorem 1.1 in \cite{Lewis}, we have the following.

\begin{theorem} Suppose $f: [0, 1] \rightarrow \IR$ is a continuously
differentiable concave function with derived function $f'(x)$.
Then the gradient function of the scalar function $A \mapsto \tr f(A)$ is
given by $f'(A) = \sum f'(a_j) P_j$ if $A$ has spectral
decomposition $A = \sum a_j P_j$.
\end{theorem}

Applying the result to the von Neumann entropy and R\'{e}nyi entropy, we have

\begin{corollary} The gradient of the objective function $S(\rho)=-\tr (\rho\log \rho)$
is
\begin{equation}
 \nabla S(\rho)=-\log\rho -I_{n_1n_2}.
\end{equation}
The gradient of the objective function $S_\alpha(\rho) =
S_\alpha(\rho) =\frac{1}{1-\alpha}\log\tr(\rho^\alpha) = \frac{1}{1-\alpha}
\log\left(\sum \lambda_j^{\alpha} \right)$ is
\begin{equation} \nabla S_\alpha(\rho) = (\tr \rho^\alpha)^{-1} \alpha \rho^{\alpha -1}.
\end{equation}
\end{corollary}

\iffalse

\it Proof. \rm According to the definition of the matrix logarithm function, we have
$$\begin{array}{lll} f(\rho)&=& \tr (\rho\log \rho) \\
&=&  \tr[\sum\limits_{n=1}^{\infty}\frac{(-1)^{n-1}}{n}(\sum\limits_{r=0}^{n}(-1)^rC_{n}^{r}\rho^{n-r+1})]  \\
&= & \sum\limits_{n=1}^{\infty}\frac{(-1)^{n-1}}{n}(\sum\limits_{r=0}^{n}(-1)^rC_{n}^{r}\tr(\rho^{n-r+1})).
\end{array}$$
Hence the gradient of the objective function is $$\begin{array}{lll} \nabla f(\rho)
&=&  \sum\limits_{n=1}^{\infty}\frac{(-1)^{n-1}}{n}(\sum\limits_{r=0}^{n}(-1)^rC_{n}^{r}\rho^{n-r})  \\
&= & ln\rho +I_{mn}.
\end{array} $$ The proof is completed.
\fi

In the following, we compute the projection operator $\Phi_{\Omega_{1}\cap \Omega_{3}}(Z).$
There is no analytic expression
of $\Phi_{\Omega_{1}\cap \Omega_{3}}(Z).$
Fortunately, we can use the Dykstra's algorithm to derive it,
which can be stated in Algorithm \ref{2.6}.
The following lemma is useful;
\iffalse
We begin with a lemma. For any $Z\in C^{n_1n_2\times n_1n_2},$ 
$E=\frac{Z^{*}+Z}{2}$ is a Hermitian matrix, and we can find the spectral
decomposition of the matrix $E$: 
$$E=U\Lambda U^{*},$$
where
$UU^{*}=I_{n_1n_2}$ and $\Lambda=\diag(\lambda_{1}(E)), \lambda_{2}(E)),
\cdots, \lambda_{n_1n_2}(E)).$
Then we have the following; 
\fi see for example, \cite[Theorem 2.1]{Higham}.

\begin{lemma} \label{2.5} Let $Z \in H_{n_1n_2}$ with spectral
decomposition 
$U\diag(\lambda_{1}(Z),  \cdots, \lambda_{n_1n_2}(Z))U^{*}$, 
where $U$ is unitary. The  projection of $Z$
onto $\Omega_{3}$ is
\begin{equation}
 \Phi_{\Omega_{3}}(Z)=U\diag(t_{1}, t_{2}, \cdots, t_{n_1n_2})U^{*}, 
\end{equation}
where $$t_{i}=\left \{ \begin{array}{ll}
\lambda_{i}(Z),\ \ & \lambda_{i}(Z)\geq 0; \\ 0, \ \ & \lambda_{i}(Z)< 0.
\end{array}\right. $$
\end{lemma}

In the following Dykstra's algorithm, the projection operator $\Phi_{\Omega_{1}}(Z)$
is defined by
Theorem \ref{2.1} and the projection operator $\Phi_{\Omega_{3}}(Z)$ is defined by
Lemma \ref{2.5}.\bigskip

\noindent\fbox{
\begin{minipage}{0.98\textwidth}
\begin{algorithm} \label{2.6}
{\sc Alternating Projection Scheme to find $\rho=\Phi_{\Omega_{1}\cap \Omega_{3}}(Z)$}\bigskip\\
\begin{tabular}{ll}
{\bf Step 1.} & Choose a positive integer $N$\! (iteration limit) and a small positive $\delta$\! (tolerance).\!\\
& Set $X_{2}^{(0)}=Z$ and do the following steps for $k=1,2,\ldots,N$. 
\medskip\\
{\bf Step 2.} & Let $X_1^{(k)}=\Phi_{\Omega_1}(X_2^{(k-1)})$ and $X_2^{(k)}=\Phi_{\Omega_3}(X_1^{(k)})$. \medskip\\
{\bf Step 3.} & If $|| X_1^{(k)}-X_2^{(k)}||<\delta$, then stop and declare $X_2^{(k)}$ a solution. \\
& Otherwise repeat step 2 until a solution is found or until $k=N$.
\end{tabular}
\end{algorithm}
\end{minipage}}\ \\

By \cite{Boyle-Dykstra}, 
one can show that the matrix sequences $\{X_{1}^{(k)}\}$ and $\{X_{2}^{(k)}\}$ generated
by Algorithm \ref{2.6} converge to the projection $\Phi_{\Omega_{1}\cap \Omega_{3}}(Z),$  that is
$$X_{1}^{(k)}\rightarrow\Phi_{\Omega_{1}\cap \Omega_{3}}(Z),\ \
X_{2}^{(k)}\rightarrow \Phi_{\Omega_{1}\cap \Omega_{3}}(Z), \ \ k\rightarrow +\infty.$$
Thus, Algorithm \ref{2.6} will determine the projection operator $\Phi_{\Omega_{1}\cap \Omega_{3}}(Z).$

Next, we use the nonmonotone spectral projected gradient method (see \cite{Birgin1,Birgin2}
for more details) to solve the minimization problem (\ref{minent}).
The algorithm starts with $\rho_{0}\in \Omega_1\cap\Omega_3$ and
use an integer $M\geq 1;$ a small parameter $\alpha_{\min}>0;$ a large parameter
$\alpha_{\max}>\alpha_{\min};$
a sufficient decrease parameter $r\in(0,1)$ and safeguarding parameters
$0<\sigma_{1}<\sigma_{2}<1.$
Initially, $\alpha_{0}\in[\alpha_{\min}, \alpha_{\max}]$ is arbitrary. Given $\rho_{t}\in \Omega$
and $\alpha_{t}\in [\alpha_{\min}, \alpha_{\max}],$ Algorithm \ref{2.8} describes how to obtain
$\rho_{t+1}$
and $\alpha_{t+1},$ and when to terminate the process. In the following algorithm, the gradient
$\nabla f(\rho)$ is defined in Lemma \ref{2.5} and the projection operator
$\Phi_{\Omega_{1}\cap \Omega_{3}}(\cdot)$
is computed by Algorithm \ref{2.6}.\bigskip

\noindent\fbox{
\begin{minipage}{0.98\textwidth}
\begin{algorithm} \label{2.8}
{\sc Scheme to solve Problem (\ref{minent})} \bigskip\\
\begin{tabular}{ll}
{\bf Step 1.} & Let $\delta$ be a small positive number. Detect whether the current point is stationary.\\
& If $\|\Phi_{\Omega_{1}\cap \Omega_{3}}(\rho_t-\nabla f(\rho_t))-\rho_t\|_{F}\leq \delta,$ then stop and declare $\rho_{t}$ a stationary point.\medskip\\
{\bf Step 2.} & Backtracking \medskip\\
{\bf Step 2.1.} & Compute $d_{t}=\Phi_{\Omega_{1}\cap \Omega_{3}}(\rho_{t}-\alpha_{t}\nabla f(\rho_t))-\rho_t$.
Set $\lambda\leftarrow 1.$ \medskip\\
{\bf Step 2.2.} & Set $\rho_+=\rho_t+\lambda d_t.$ \medskip\\
{\bf Step 2.3.} & If $ f(\rho_+)\leq \max\limits_{0\leq j \leq \min\{t, M-1\}}f(\rho_{t-j})+
\gamma\lambda\langle d_t,\nabla f(\rho_t)\rangle,$ then define \medskip\\
& $\lambda_t=\lambda, \ \rho_{t+1}=\rho_+,\ s_t=\rho_{t+1}-\rho_{t},\ y_t=\nabla f(\rho_{t+1})-\nabla f(\rho_{t}),$
and go to Step 3.\medskip\\
& 
Otherwise, define $\lambda_{new}=\frac{\sigma_{1}\lambda
+\sigma_{2}\lambda}{2} \in [\sigma_1\lambda,\sigma_2\lambda],$ set $\lambda\leftarrow\lambda_{new}$, and go to Step 2.2.\medskip\\
{\bf Step 3.} & Compute $b_t=\langle s_t, y_t\rangle.$ If $b_t\leq 0$, set 
$\alpha_{t+1}=\alpha_{\max},$
else, compute $\alpha_{t}=\langle s_t, s_t\rangle$\\
&  and
 $\alpha_{t+1}= \min\{\alpha_{\max},\max\{\alpha_{\min}, \frac{a_t}{b_{t}}\}\}.$
\end{tabular}
 \end{algorithm}
\end{minipage}}\ \\

By Theorem 2.2 in \cite{Lewis}, the sequence $\{\rho_t\}$ generated by Algorithm \ref{2.8} 
converges to the solution of the minimization problem {\rm (\ref{minent})}.

A computational comment can be made on Algorithm \ref{2.8}.
In order to guarantee the iterative sequence 
${\rho_{t}}\in \Omega_{1}\cap \Omega_{3}, t=0,1,2,\cdots,$ the
initial value $\rho_{0}$ must be in $\Omega_{1}\cap \Omega_{3}.$ Taking $\rho_{1}$
for example, if $\rho_{0}\in \Omega_{1}\cap \Omega_{3},$ then $\rho_{1}=\rho_{0}+\alpha_{1}d_{1}
\in \Omega_{1}\cap \Omega_{3},$ because $d_{1}=\Phi_{\Omega_{1}\cap \Omega_{3}}
(\rho_{0}-t_{0}\nabla f(\rho_{0}))-x_{0}\in \Omega_{1}\cap \Omega_{3}$ and $\alpha_{1}$ is a scalar.

\section{Multipartite States} 
In this section, we will use projection methods to find a global state in a multipartite system
with prescribed reduced states.
That is, letting $\emptyset\neq J_1,\ldots,J_m \subset \underline{\mathbf{k}}$ denote the indices of a given family of subsystems of a $k$-partite system on
$H_{n_1}\otimes \ldots \otimes H_{n_k}$, can we find a global quantum state $\rho\in D_{n_1\cdots n_k}$
with prescribed reduced states
\[\tr_{J_1^c}(\rho)=\rho_{J_1},\quad \tr_{J_2^c}(\rho)=\rho_{J_2},\quad  \ldots,\quad
\tr_{J_m^c}(\rho)=\rho_{J_m}?\]
For example, if $k = 3$, one may need to find a global state $\rho \in D_{n_1n_2n_3}$ 
with prescribed reduced states:
$\tr_1(\rho) = \rho_{23} \in D_{n_2n_3}$ and $\tr_3(\rho) = \rho_{12} \in D_{n_1n_2}$.
We will further require the global state $\rho$ to have prescribed eigenvalues.

We will extend the results in the previous section to multipartite systems.
Note that the study is more challenging. For example, to find a global sate 
$\rho \in D_{n_1\cdot n_2}$ with prescribed states $\tr_2(\rho) = \rho_1$ and
$\tr_1(\rho) = \rho_2$, one can replace 
$(\rho, \rho_1, \rho_2)$ by $((U\otimes V)^*\rho(U\otimes V), U^*\rho_1U, V^* \rho_2V)$
for some suitable unitary $U \in M_{n_1}$ and $V \in M_{n_2}$
and assume that $\rho_1, \rho_2$ are in diagonal form.
However, to find $\rho\in D_{n_1n_2n_3}$ with prescribed reduced states
$\tr_1(\rho) = \rho_{23}$ and $\tr_3(\rho) = \rho_{12}$, there is no easy transform 
to reduce the problem to the case when $\rho_{12}$ and $\rho_{23}$ are in diagonal form.

To use the projection methods, we need to find the 
least square projection of
a hermitian matrix $Z \in H_{n_1 \cdots n_k}$
to the linear manifold
\begin{equation}\label{multi}
\mathcal{L} = \{X\in H_{n_1 \cdots n_k} : \tr_{J_i^c}(X) = \rho_{J_i}, i = 1, \dots, m\}. 
\end{equation}
In the following proposition, we answer this problem for $m=1$. (See Appendix \ref{proofs6} for the proof.) 

\begin{proposition}\label{multiprob} Let $J\subseteq \underline{\mathbf{k}}$. Given $Z\in H_{n_1 \cdots n_k}$,
the least square approximation of $Z$ in the linear manifold $\mathcal{L}=\{\rho\in H_{n_1 \cdots n_k}:
\tr_{J^c}(\rho)=\sigma\}$ is given by
\begin{equation}
\Phi_{\mathcal{L}}(Z)=Z-\mathcal{M}_{J}(Z,\sigma),
\end{equation}
where
\begin{equation}\label{pseudo}
\mathcal{M}_{J}(Z,\sigma)=P_J^{T}\left(\frac{I_{n_{J^c}}}{n_{J^c}}\otimes
(\tr_{J^c}(Z)-\sigma)\right)P_J,
\end{equation}
$n_{J^c}=\prod\limits_{i\in J^c}^k n_i$ and $P_J$ is the permutation matrix such that
\begin{equation}\label{permute}
P_J(\alpha_1\otimes \alpha_2\otimes \cdots \otimes \alpha_k)P_J^{T}
=\bigotimes\limits_{i\in J^c} \alpha_i
\otimes \bigotimes\limits_{i\in J} \alpha_i.
\end{equation}
\end{proposition}
%Next, we consider a set of linear constraints with overlaps. That is, 
%\[L=\{x\in \mathbb{R}^{N}: A_{j}x=b_{j}\mbox{ for all } j=1,\ldots,r\}\]
%where $Row(A_{i})\cap Row(A_j)\neq \emptyset$ for some $i\neq j$. If there are only two overlapping constraints, we can compute the least square approximation of $x\in \mathbb{R}^{N}$ in $L$ as follows. 
%
%\begin{proposition} \label{3.2}
%Let $A_1\in M_{n_1,N}$ and $A_2\in M_{n_2,N}$ and $P$ and $Q$ be orthogonal projections such that $A_0=PA_1=QA_2$, where $Row(A_0)=Row(A_1)\cap Row(A_2)$. Then the set 
%\[L=\{x\in \mathbb{R}^{N}: A_{1}x=b_{1} \mbox{ and } A_2x=b_2\}\neq \emptyset\] if and only if $Pb_1=Qb_2=b$. 
%In this case, the least square approximation of $x$ in $L$
%equals
%$$\tilde x = x - A_1^+(A_1x-b_1) - A_2^+(A_2x-b_2) + A_{0}^+(A_0 x - b).$$
%\end{proposition}
We now extend the formula given in equation (\ref{pseudo}) to the general case. To familiarize the reader with the notation in the next proposition, let us start with an example. Let $k=3=m$ and $J_1=\{1,2\}$, $J_2=\{23\}$ and $J_3=\{3\}$. Given $\rho_{J_1}\in D_{n_1n_2}$, $\rho_{J_2}\in D_{n_2n_3}$ and $\rho_{J_3}\in D_{n_3}$, then the set $\mathcal{L}$ defined in equation (\ref{multi}) is $\mathcal{L}=\{X\in H_{n_1n_2n_3}\ : \ \mbox{tr}_{J_1^c}(X)=\rho_{J_1},\mbox{tr}_{J_2^c}(X)=\rho_{J_2}, \mbox{ and } \mbox{tr}_{J_3^c}(X)=\rho_{J_3}\}$. Note that $J_1\cap J_2=\{2\}$ so that if $\mathcal{L}$ contains an element $X$, then it must hold that
 \[\mbox{tr}_{13}(X)=\mbox{tr}_{(J_1\cap J_2)^c}(X)=\mbox{tr}_{1}(\rho_{J_1})=\mbox{tr}_3(\rho_{J_2}):=\rho_{J_1\cap J_2}=\rho_2.\]
Similarly, since $J_2\cap J_3=\{3\}$, we must have  
 \[\mbox{tr}_{12}(X)=\mbox{tr}_{(J_2\cap J_3)^c}(X)=\mbox{tr}_{3}(\rho_{J_2})=\rho_{J_3}.\]
Note that since $\rho_{J_1},\rho_{J_2}$ and $\rho_{J_3}$ are density matrices, then it also follows that 
\[\mbox{tr}_{123}(X)=\mbox{tr}(X)=\mbox{tr}_{J_1}(\rho_{J_1})=\mbox{tr}_{J_2}(\rho_{J_2})=\mbox{tr}_{J_3}(\rho_{J_3})=1\]
In fact, the first two conditions above, namely $\mbox{tr}_1(\rho_{J_1})=\mbox{tr}_3(\rho_{J_2})$ and $\mbox{tr}_3(\rho_{J_2})=\rho_{J_3}$ are enough to guarantee that $\mathcal{L}$ is non-empty. This is stated in the following proposition, whose proof can be found in Appendix \ref{proofs6}.

\begin{proposition}\label{multiprojform}
Let $J_1,\ldots, J_m\subseteq \underline{\mathbf{k}}$ and  $\mathcal{L}$ be defined as in (\ref{multi}). Then $\mathcal{L}\neq \emptyset$ if and only if for any $\mathcal{S} \subseteq \{J_1,\ldots, J_m\}$ and any $T_1,T_2\in \mathcal{S}$, 
\[\tr_{\tilde{T}_1}(\rho_{T_1})=\tr_{\tilde{T}_2}(\rho_{T_2}):=\rho_{S_{int}}, \mbox{ where } 
\mathcal{S}_{int}=\bigcap\limits_{T\in \mathcal{S}}T \mbox{ and }\tilde{T}_j=T_j\setminus \mathcal{S}_{int} \mbox{ for } j=1,2 \]
Furthermore, the least square approximation of a given $Z\in H_{n_1 \cdots n_k}$ is 
\begin{equation}
\Phi_{\mathcal{L}}(Z)=Z+\sum\limits_{\emptyset\neq\mathcal{S}\subseteq \{J_{1},\ldots,J_m\}}(-1)^{|\mathcal{S}|}
 \mathcal{M}_{\mathcal{S}_{int}}\left(Z,\rho_{\mathcal{S}_{int}}\right), 
\end{equation}
where $\mathcal{M}_{\mathcal{S}_{int}}\left(Z,\rho_{\mathcal{S}_{int}}\right)$ is as defined in equation 
(\ref{pseudo}). 
\end{proposition}

Suppose we are interested in looking for a tripartite state $\rho\in \mathcal{D}_{n_1n_2n_3}$ with given partial traces $\mbox{tr}_1(\rho)=\rho_{23}$ and $\mbox{tr}_3(\rho)=\rho_{12}$. Then we can use Proposition \ref{multiprojform} to obtain the following projection formula.  
\begin{corollary}
%Let $T_1:H_{n_1n_2n_3}\longrightarrow H_{n_2n_3}$,
%$T_2:H_{n_1n_2n_3}\longrightarrow H_{n_1n_2}$ and $T_3:H_{n_1n_2n_3}\longrightarrow H_{n_2}$
%such that
%$T_1(\rho)=\mbox{tr}_1(\rho),T_2(\rho)=\mbox{tr}_3(\rho),$ and $T_3(\rho)=\mbox{tr}_{13}(\rho)$.
Suppose $\rho_{12}\in D_{n_1n_2}$ and $\rho_{23}\in D_{n_2n_3}$. The set
\begin{equation} \label{6.3L}
\mathcal{L}=\{X\in H_{n_1n_2n_3}: \tr_1(\rho)=\rho_{23} \mbox{ and } \tr_3(\rho)=\rho_{12}\}
\end{equation}
is nonempty if and only if $\tr_{1}(\rho_{12})=\gamma=\tr_{3}(\rho_{23})$. In this case, the least square approximation of a given $Z\in H_{n_1n_2n_3}$ onto the set $\mathcal{L}$ in (\ref{6.3L})  is given by
\[\begin{array}{lcl}
\Phi_{\cL}(Z)& = & Z-\left[\frac{I_{n_1}}{n_1}\otimes (\tr_1(Z)-\rho_{23})\right]-\left[(\tr_3(Z)-\rho_{12})\otimes \frac{I_{n_3}}{n_3}\right]+\left[\frac{I_{n_1}}{n_1}\otimes (\tr_{13}(Z)-\gamma)\otimes \frac{I_{n_3}}{n_3}\right]
\end{array}
\]
\end{corollary}
We employ the following alternating projection method to determine if there exists
$\rho\in \Omega_3\cap \mathcal{L}$, where \[\Omega_3=\{\rho\in M_{N} \ : \ \rho\geq 0\}
\mbox{ and } \mathcal{L}=\{\rho \ : \ \tr_{J_1^c}=\sigma_{J_1}, \ldots, \tr_{J_m^c}=\sigma_{J_m}\}.\]
The following algorithm is a generalization of Algorithm \ref{2.3} and \ref{2.6} to multipartite systems. One must first check that $\mathcal{L}\neq \emptyset$ using Proposition \ref{multiprojform}. We will use $\Phi_{\mathcal{L}}$ and $\Phi_{\Omega_3}$ as defined by Proposition \ref{multiprojform} and Lemma \ref{2.5}.\medskip

\noindent\fbox{
\begin{minipage}{0.98\textwidth}
\begin{algorithm}{\sc Construction of a state $\rho\in \Omega_i\cap \mathcal{L}$, where $i=2$ or $i=3$.}\ \label{3.8}\bigskip\\
\begin{tabular}{ll}
{\bf Step 1.} & Choose a positive integer $N$ (say $N=1000$) as iteration limit and a small positive \\
& number $\delta$ (say $\delta=10^{-15}$) as a error/tolerance value and set $k=0$. \\
{\bf Step 2.} & Generate a random unitary $U$ and a random probability vector
$(d_1,\ldots,d_{n_1\ldots n_k})$ and \\
& set the initial point to be $\rho_0=U
\diag(d_1,\ldots, d_{n_1\ldots n_k})U^*$. Do the next step for $k\leq N$. \\
{\bf Step 3.} & For $k\geq 1$, let $\rho_{2k-1}=\Phi_{\mathcal{L}}(\rho_{2k-2})$ and $\rho_{2k}=\Phi_{\Omega_i}(\rho_{2k-1})$ as defined by Proposition \ref{multiprojform}\\
&  and Lemma \ref{2.5} or Theorem \ref{2.1}. If $||\tr_1(\rho)_{2k}-\rho_2||+||\tr_2(\rho)_{2k}-\rho_1||<\delta$, then stop\\
&  and declare $\rho_{2k}$ as a solution. 
\end{tabular}
\end{algorithm}
\end{minipage}}

\section{Numerical Experiments }\label{numex} \vskip2mm

In this section, some examples are tested to illustrate that Algorithms \ref{2.3}, \ref{altprojrank}, \ref{algrankk}, \ref{alginterlace}, \ref{lowrankC} and \ref{3.8}  are feasible and effective to solve Problem 1.1. All experiments are performed in MATLAB R$_{2015a}$
on a PC with an Intel Core i7 processor at 2.40GHz with machine precision
$\varepsilon=2.22\times 10^{-16}.$
The programs can be downloaded from \verb|http://cklixx.people.wm.edu/mathlib/projection/|.

\subsection{Algorithm \ref{2.3} for solving Problem 1.1 with the prescribed eigenvalues}\vskip 2mm
In this subsection, we present a simple numerical example to illustrate that Algorithm \ref{2.3} is feasible
to solve Problem 1.1 with the prescribed eigenvalues.
In Algorithm \ref{2.3}, note that $X_{2k}\in \Omega_2$, that is, $X_{2k}$ has the prescribed eigenvalues $c_1,\ldots,c_{n_1n_2}$. Now, define $$Err(X_{2k})=\|\mbox{tr}_1(X_{2k})-\rho_{2}\|+\|\mbox{tr}_2(X_{2k})-\rho_{1}\|,$$
Hence, $X_{2k}\in \Omega_1\cap \Omega_2$ if and only if $Err(X_{k})=0$. When implementing Algorithm \ref{2.3}, we declare $X_{2k}$ a solution if $Err(X_{2k})<\delta$ for some small positive number $\delta$. If this criteria is not met after a set number of iterations, then the algorithm terminates. 

\begin{example}\label{ex1} Set $(c_1,c_2,c_3,c_4,c_5,c_6)=(0.8329,0.0781,0.0529,0.0238,0.0109,0.0015)$  and 
%$$\rho_{2}=\left(\begin{array}{ccc}
%0.3282  &  0.2362  &  0.2309 \\
%0.2362  &  0.2920   & 0.2414\\
%0.2309   & 0.2414  &  0.3798
%\end{array}\right),\ \
%\rho_{1}=\left(\begin{array}{cc}
%0.5523  &  0.3984 \\
%0.3984  &  0.4477\\
%\end{array}\right).$$
\[\rho_2=\begin{pmatrix}
0.4922 & 0.2729 & 0.3138 \\
0.2729 & 0.1980 & 0.1846\\
0.3138 & 0.1846 & 0.3098
\end{pmatrix}, \quad \rho_1=\begin{pmatrix}
0.52 & 0.3923\\
0.3923 & 0.48
\end{pmatrix}. \]
Using Algorithm  \ref{2.3}, we obtain the following solution after $214$ iterations and $Err(X_{214})\approx 3.38\times 10^{-16}.$ 
$$X\approx X_{214}=\left(\begin{array}{cc}
X_{11}^{(214)} & X_{12}^{(214)} \\
X_{21}^{(214)} & X_{22}^{(214)}\\
\end{array}\right)=\left(\begin{array}{cccccc}
    0.2826  &  0.1614 &   0.1582  &  0.1990 &   0.0908 &  0.1861\\
   0.1614 &  0.1234  &  0.0945  &  0.1258 &   0.0601 &  0.1234 \\
   0.1582  &  0.0945  &  0.1140 &   0.1088  &  0.0470 &  0.1333 \\
    0.1990  & 0.1258  &  0.1088  &  0.2096   & 0.1115 & 0.1556  \\
    0.0908  &  0.0601 &  0.0470  & 0.1115  &  0.0746 &  0.0901  \\
  0.1861  &  0.1234 &  0.1333  &  0.1556  &  0.0901 &  0.1958
\end{array}\right),$$
\end{example}

Example \ref{ex1} illustrates the effectiveness of Algorithm \ref{2.3} in solving Problem 1.1 with prescribed eigenvalues.
\subsection{Algorithms \ref{altprojrank}, \ref{algrankk} , \ref{alginterlace} and \ref{lowrankC} to find solutions with prescribed rank}  In Subsection 2.2, we discussed four different algorithms to find a low rank solution $\rho\in \cS(\rho_1,\rho_2)$. 

Let $r_1=\rank(\rho_1)$ and $r_2=\rank(\rho_2)$ and $r=\rank(\rho)$. Also, let  \[err=\max\{||\rho_1-\mbox{tr}_2(\rho)||,||\rho_2-\mbox{tr}_1(\rho)||\}.\] Denote by $\lambda_M$ and $\lambda_{\mu}$ the maximum and minimum eigenvalues of $\rho$, respectively; and $ent$ the Von Neumman entropy of $\rho$. The following table illustrates the performance of each algorithm.

\begin{example}\label{ex2}
 Consider $\rho_1\in D_3$ and $\rho_2\in D_4$ with eigenvalues \[\lambda(\rho_1)=(0.5951,0.2341,0.1708)\quad  \lambda(\rho_2)=(0.6124,0.1926,0.1654,0.0296)\]
\begin{tabular}{|l|c|c|c|c|c|c|c|c|c|}
\hline & $(m,n)$ & $(r_1,r_2)$ & $r$ & CPU-time & err & $\lambda_M$ &$\lambda_{\mu}$ & ent\\ 
 \hline  Alg\ref{algrankk} & (3,4) & (3,4) & 4 & 0.002s  & 3.54294e-17 & 0.399619 &  -6.00329e-17 & 1.27929\\
 \hline  Alg\ref{alginterlace} & (3,4) & (3,4) & 3 & 0.006s & 1.11022e-16 &  0.9313 & -1.48157e-16  & 0.297223\\
 \hline  Alg\ref{lowrankC} & (3,4) & (3,4) & 3 & 0.004s & 1.11022e-16  & 0.9531 & -4.1612e-17 &  0.215848\\
 \hline 
\end{tabular}\medskip\\
Using Algorithm \ref{altprojrank}, we determine if we can find a solution of rank $2,\ldots, \rank(X_0)-1$, where $X_0$ is a solution obtained from one of the algorithms above. Here are the solutions we obtained. \medskip\\
\begin{tabular}{|c|c|c|c|c|c|c|c|c|c|c|}
\hline  $(m,n)$ & $(r_1,r_2)$ & $X_0$ & $r$ & $\#$ iter & CPU-time & err & $\lambda_M$ & $\lambda_{\mu}$ & ent\\ 
 \hline  (3,4) & (3,4) & alg. \ref{alginterlace} & 2 & 1336 & 0.54s & 9.34747e-16       &  0.9017 & -4.16498e-17 &   0.321332 \\
 \hline  (3,4) & (3,4) & alg. \ref{lowrankC} & 2 & 3103 & 1.266s & 9.85657e-16 & 0.9531 & -5.19103e-17             &  0.189284\\
 \hline  
\end{tabular}\bigskip\\
Note that in this case, the solution obtained by Algorithm \ref{altprojrank} using the solution from Algorithm \ref{lowrankC} as initial point, has minimum entropy in $\mathcal{S}(\rho_1,\rho_2)$. This is because $\rho$ is rank $2$ and the largest eigenvalue of $\rho$ is the maximum possible eigenvalue of any element of $\mathcal{S}(\rho_1,\rho_2)$. 
\end{example}

\begin{example}\label{ex3}
Consider $\rho_1\in D_3,\rho_2\in D_6$ such that  
\[\lambda(\rho_1)=(0.8213,0.1234,0.0553) \mbox{ and } \lambda(\rho_2)= (0.5720,0.3068,0.1000,0.0189,0.0020,0.0003)\]
\begin{tabular}{|l|c|c|c|c|c|c|c|c|c|}
\hline Algorithm & $(m,n)$ & $(r_1,r_2)$ & $r$ & CPU-time & err & $\lambda_M$ &$\lambda_{\mu}$ & ent\\ 
 \hline  \ref{algrankk} & (3,6) & (3,6) & 6 & 0.003s & 8.9182e-16 &  0.469983   &-4.93499e-17 &     1.19924 \\
 \hline  \ref{alginterlace}&(3,6) & (3,6) & 4 &  0.005s & 3.31468e-16 & 0.690947 &-6.27654e-17 & 0.632879\\
 \hline  \ref{lowrankC} & (3,6) & (3,6) &  6&   0.004s & 2.78333e-16 &    0.750675   & -5.4791e-17    &    0.755308\\
 \hline 
\end{tabular}
\begin{center}
Algorithm \ref{altprojrank}
\end{center}
\begin{tabular}{|c|c|c|c|c|c|c|c|c|c|c|}
\hline  $(m,n)$ & $(r_1,r_2)$ & $X_0$ & $r$ & $\#$ iter & CPU-time & err1 & $\lambda_M$ & $\lambda_{\mu}$ & ent\\ 
 \hline  (3,6) & (3,6) & alg. \ref{lowrankC} & 3 & 76933 & 44.25s &  9.90465e-16 &    0.729479  & -5.79165e-17       &  0.736448 \\
 \hline  (3,6) & (3,6) & alg. \ref{alginterlace} & 2  & 100000  & 63.5203s  & 2.26889e-08   & 0.690947  & -1.44764e-16 & 0.618341   \\
 \hline  (3,6) & (3,6) & alg. \ref{alginterlace} & 3  & 6707 & 4.39s  & 9.83117e-16  & 0.690947 & -6.84736e-17       &  0.631907 \\
 \hline  
\end{tabular}
\end{example}

\begin{example}\label{ex4}
 In this example, we consider $\rho_1\in D_6,\rho_2\in D_8$ such that  
\[\lambda(\rho_1)=(0.2272,0.2136,0.1946,0.1474,0.1341,0.0831)\] 
\[\mbox{ and }\quad \lambda(\rho_2)= (0.2399,0.1699,0.1638,0.1463,0.1246,0.0851,0.0407,0.0297)\]
\begin{tabular}{|l|c|c|c|c|c|c|c|c|c|}
\hline Algorithm & $(m,n)$ & $(r_1,r_2)$ & $r$ & CPU-time & err & $\lambda_M$ &$\lambda_{\mu}$ & ent\\ 
 \hline  \ref{algrankk} & (6,8) & (6,8) & 8 & 0.005s & 2.56989e-16 & 0.151124  & -3.91005e-17        &       2.0642 \\
 \hline  \ref{alginterlace}& (6,8) & (6,8) & 3 & 0.014s & 4.38087e-16 & 0.840737 & -1.36117e-16      &  0.515135  \\
 \hline  \ref{lowrankC} & (6,8) & (6,8) & 4 & 0.017s & 3.08212e-16 &  0.914875 & -1.05048e-16        & 0.308127 \\
 \hline 
\end{tabular}
\begin{center}
Algorithm \ref{altprojrank}
\end{center}
\begin{tabular}{|c|c|c|c|c|c|c|c|c|c|c|}
\hline  $(m,n)$ & $(r_1,r_2)$ & $X_0$ & $r$ & $\#$ iter & CPU-time & err1 & $\lambda_M$ & $\lambda_{\mu}$ & ent\\ 
 \hline  (6,8) & (6,8) & alg. \ref{lowrankC} & 3 & 26770 & 45.955s & 8.97652e-16  & 0.914681 & -8.7338e-17             &  0.308847 \\
 \hline  
\end{tabular}
\end{example}

\subsection{Algorithm \ref{3.8} for solving Problem 1.1 with multipartite states}

In this section, we give two examples using Algorithm \ref{3.8} to find a global state in a multipartite system with prescribed reduced states and prescribed eigenvalues.

%\vskip .5in
%\section{Numerical examples,  computational issues, and related problems}
%\begin{example} \rm
\begin{example}\label{ex5}
We implement Algorithm \ref{3.8}, for $i=3$, to find a tripartite state $\rho\in D(2\cdot  2 \cdot 2)$ such that $\tr_1(\rho)=\rho_1$ and $\tr_3(\rho)=\rho_2$, where
\[\rho_1=
\begin{pmatrix} 0.181375 &	0.161 &	0.1678 &	0.1417\\
0.161	& 0.314875 &	0.2653	& 0.1937\\
0.1678	& 0.2653	& 0.307275 &	0.1863\\
0.1417	& 0.1937	& 0.1863	& 0.196475\end{pmatrix}\in D(2^2),\]
\[\rho_2=\begin{pmatrix}
0.214875 &	0.1653 &	0.1926	& 0.1934\\
0.1653 &	0.264475 &	0.2166	& 0.1888\\
0.1926 &	0.2166 &	0.281375 &	0.1962\\
0.1934 &	0.1888	& 0.1962 &	0.239275\end{pmatrix}\in D(2^2).\]
%\end{example}
Using a random $8\times 8$ Hermitian matrix as the initial point for the projection scheme, the algorithm produced the solution \small
\[\rho=\begin{pmatrix}
0.0811 & 	0.0809 & 	0.0747 & 0.0654 &	
0.0850 &	0.0901 & 	0.0923 & 0.07\\
0.0809 &	0.1338 &	0.1189 & 0.0906 &	
0.0898 & 	0.1076 &	0.1003 & 0.1011 \\
0.0747 & 	0.1189 &	0.1637 & 0.0893 &	
0.1053 &	0.0658 &	0.0944 & 0.0947\\
0.0654 &	0.0906 &	0.0893  &	0.1008 &	0.0728 &	0.1113 &	0.1013 &	0.0944 \\
0.085 &	0.0898 &	0.1053 &	0.0728 & 	0.1003 &	0.0801 &	0.0931 &	0.0763 \\
0.0901 &	0.1076 &	0.0658 &	0.1113 &	0.0801 &	0.1811 &	0.1464 &   0.1031 \\
0.0923 &	0.1003 &	0.0944 &	0.1013 &	0.0931 &	0.1464 &	0.1436 &   0.097 \\
0.07 &	0.1011 &	0.0947 &	0.0944 &	0.0763 &	0.1031 &	0.097 &	0.0957
\end{pmatrix}\]\normalsize
with an error of order not more than $10^{-16}$. This rank 6 solution is found after approximately 400 iterations, where one iteration consists of a projection on $\Phi_3$ and a projection on $\Phi_{\mathcal{L}}$. The result was obtained in approximately 0.3 seconds. 
Note that if $n_1=n_3=2$ and $n_2$ is increased to $n=8$, this program still obtains a solution relatively fast and accurately.
\end{example}

\begin{example}\label{ex6}
We implement Algorithm \ref{3.8}, for $i=2$, to find
$\rho\in D_8$ with $\tr_1(\rho)=\rho_1$, $\tr_3(\rho)=\rho_2$ (as in the previous example) with the additional condition that the eigenvalues of $\rho$
are \[\lambda(\rho)=(0.8034, 0.0889, 0.05204,
0.0284, 0.0188, 0.0051,
0.0032, 0.0001).\]
%\end{example}
The algorithm ran in under 0.2 seconds and approximately 300 iterations to
produce the solution
\[\rho=\begin{pmatrix}
0.1507  &  0.1056 &   0.0999  &  0.0769 & 0.1047  & 0.0966  &  0.1264   &  0.1293\\
0.1056  &  0.1209 &   0.0977  &  0.0716 &   0.0813  &  0.0792 &   0.1248  &  0.1018\\
0.0999  &  0.0977 &  0.1144   & 0.0680 & 0.0879  &  0.0685 &   0.1241  &  0.1100\\
0.0769  & 0.0716  &  0.0680   & 0.1274 & 0.1053  & 0.0559  &  0.0836   & 0.0821 \\
0.1047  & 0.0813  &  0.0879   & 0.1053 & 0.1160  &  0.0818 &  0.0990   & 0.1055\\
0.0966  &  0.0792 &   0.0685  &  0.0559 & 0.0818  &  0.0832 &   0.0795  &  0.0870\\
0.1264  &  0.1248 &   0.1241  &  0.0836 & 0.0990  &  0.0795 &   0.1549  &  0.1297\\
0.1293  &  0.1018 &   0.1100  &  0.0821 & 0.1055  &  0.0870 &   0.1297  &  0.1324
\end{pmatrix} \]
with with an error of order not more than $10^{-16}$. Note here that we used a random $8\times 8$ Hermitian matrix as the initial point for the projection scheme. Using a different initial point may change the solution obtained from the algorithm. 
\end{example}

\begin{example}\label{ex7} We illustrate Algorithm \ref{3.8} for the case that $\rho\in D(8)$ and $\tr_3(\rho)=\rho_{12}=\rho_{13}=\tr_2(\rho)$. Let 
\[\rho_{12}=\rho_{13}=\begin{pmatrix}
0.2471 & 0.1842 & 0.1738 & 0.2546\\
0.1842 & 0.2277 & 0.1386 & 0.2144\\
0.1738 & 0.1386 & 0.182 & 0.2303\\
0.2546 & 0.2144 & 0.2303 & 0.3432
\end{pmatrix}.\]
This type of problem is an example of a $2-$symmetric extension problem. In \cite{CJYZ}, the existence of a solution to such a problem was characterized using the concept of separability of quantum states. Using Algorithm \ref{3.8}, we find a solution
\[\rho= \begin{pmatrix}
0.1302 & 0.1096 & 0.1111 & 0.1071	&
0.0615 & 0.1156 & 0.1151 &	0.1470\\
0.1096 & 0.1169 & 0.1147 & 0.0731 &	
0.0554 & 0.1123 & 0.1139 & 0.1395 \\
0.1111 & 0.1147 & 0.1169 & 0.0746 &	
0.0547 & 0.1152 & 0.1123 & 0.1390\\
0.1071 & 0.0731 & 0.0746 & 0.1108 & 	0.0483 & 0.0839 & 0.0832 & 0.1021\\
0.0615 & 0.0554 & 0.0547 & 0.0483 &	
0.0322 & 0.0649 & 0.0650 & 0.0789 \\
0.1156 & 0.1123	& 0.1152 & 0.0839 &	
0.0649 & 0.1498 & 0.1427 & 0.1653\\
0.1151 & 0.1139 & 0.1123 & 0.0832 & 	0.0650 & 0.1427 & 0.1408 & 0.1641 \\
0.1470 & 0.1395 & 0.1390 & 0.1021 &	
0.0789 & 0.1653 & 0.1641 & 0.2024
\end{pmatrix}\]
with an error of order $10^{-17}$ after 2353 iterations in 1.9 seconds. 
\end{example}
%In terms of programming, it is convenient to associate a subset
%$J_j$ with a row vector. For tripartite system,
%we identify $J_i$ with the zero-one vector in $\IR^{1\times 3}$ so that
%the nonzero entries in the vectors $J_i$ corresponds to the indexes in the set $J_i$.
%Then the collection $J = \{J_1, \dots, J_k\}$
%can be encoded in a $k\times 3$ matrix. Of course, $J_j'$ corresponds to the
%row vector such that $J_j' + J_j$ is the vector of all ones.
%For example, if\small
%$J = {
%\begin{pmatrix} 1 & 1 & 0 \cr 0 & 1 & 1 \cr 1 & 0 & 0\cr\end{pmatrix}}$,
%\normalsize
%then
%$$\cS(\rho_J) = \{\rho \in D_{n_1n_2n_3}: \tr_3(\rho) = \rho_{12}, \tr_1(\rho) = \rho_{23},
%\tr_{23}(\rho) = \rho_1\}.$$

\section{Concluding remarks and further research}

In this paper, we 
use projection methods to construct (global) quantum states with
prescribed reduced (marginal) states,
and specific ranks and possibly extreme Von Neumann or Renyi entropy. 
Using convex analysis, optimization techniques on matrix manifolds,
we obtained convergent algorithms to solve the problem. Matlab programs
were written based on these algorithms, and numerical examples of low 
dimension cases were demonstrated. 
There are many problems deserving further investigations.
We mention a few of them in the following.

\begin{enumerate}
\item We have only demonstrated our algorithms with low dimension examples. It is interesting 
to improve the algorithm so that it can deal with practical problems (of large sizes).
 
\item Besides the alternating projection methods, it is interesting 
to study other schemes such as the Douglas-Rachford reflection method 
(for example, see \cite{DR,Sv,Phan}) to solve our problem.
 
%\item It is desirable to construct a pure state (rank one) $\rho$ with prescribed 
%reduced states.  However, because the set of rank 1 matrices
%in $D_{n_1 \cdots n_k}$ is not convex, it is not easy to find a convergent 
%algorithm to construct the global sate, and requires further investigation.

\item If it is impossible to find a pure state with the prescribed 
reduced sates, one might try  to construct 
a global state with minimum rank.  The set of matrices in 
$D_{n_1 \cdots n_k}$ with a fixed rank, or  bounded ranks, has 
complicated geometry. A test to determine if a solution produced has minimum rank is lacking.

\end{enumerate}

\bigskip\noindent
{\bf Acknowledgment}

\medskip\noindent
Duan was supported by the National Natural Science Foundation of China (No.11561015;  11761024; 11961012 ),  and the Natural Science Foundation of Guangxi Province (No. 2016GXNSFFA380009; 2017GXNSFBA198082; 2016GXNSFAA380074 ).  Li is an affiliate member of the Institute for Quantum Computing,
University of Waterloo.
His research was supported by the USA NSF grant DMS 1331021, the
Simons Foundation Grant 351047. Pelejo was supported by the University of the Philippines Diliman OVCRD through the PhDIA 191902 Research Grant.

%\nocite{*}
%\bibliography{projbib}
%\bibliographystyle{ieeetr}

%
%\bigskip
%(Duan)
%College of Mathematics and Computational Science,
%Guilin University of Electronic Technology, Guilin 541004, P.R. China.
%duanxuefeng@guet.edu.cn
%
%\medskip
%(Li)
%Department of Mathematics, College of William and Mary,
%Williamsburg, VA 23187, USA. ckli@math.wm.edu
%
%\medskip
%(Pelejo)
%Institute of Mathematics, College of Science, University of the Philippines Diliman, Quezon City 1101, Philippines, dcpelejo@math.upd.edu.ph
\bigskip

\appendix
\section{Proof of Proposition \ref{proj1}}\label{proofs1}
Let $\Phi_{\Omega_1}(P)=[X_{ij}]_{i,j\in \underline{\mathbf{n_1}}}$, where $X_{ij}\in M_{n_2}$ and let $\rho_{1}=[a_{ij}]_{i,j\in \underline{\mathbf{n_1}}}$. We wish to show that 
\begin{equation}\label{blocks}
X_{ij}=P_{ij}+\delta_{ij}\frac{1}{n_1}\left(\rho_2-\sum\limits_{i=1}^{n_1}P_{ii}\right)+\frac{1}{n_2}\left(a_{ij}-\tr(P_{ij})+\delta_{ij}\frac{\tr(P)-1}{n_1}\right)I_{n_2}
\end{equation}
Since $\Omega_{1}$ is a closed and convex set, by the definition of
projection operator we obtain that $\Phi_{\Omega_1}(P)$ is the unique solution of the minimization problem
$$\min\limits _{X\in \Omega_{1}}\|P-X\|^2
=\min\limits _{X\in \Omega_{1}}\sum\limits_{1\leq i\leq n_1}\|P_{ii}-X_{ii}\|^2+
\sum\limits_{1\leq i\neq j\leq n_1}\|P_{ij}-X_{ij}\|^2, \eqno(\ref{blocks}.1)$$
which is equivalent to
$$\min\limits _{{\left \{ \begin{array}{ll}
X_{11}+X_{22}+\cdots+X_{n_1,n_1}=\rho_2 \\
\tr(X_{ii})=a_{ii},1 \leq i\leq n_1
\end{array}\right.}}
\sum\limits_{1\leq i\leq n_1}\|P_{ii}-X_{ii}\|^2+
\min\limits _{\tr(X_{ij})=a_{ij}, 1\leq i\neq j\leq n_1}
\sum\limits_{1\leq i\neq j\leq n_1}\|P_{ij}-X_{ij}\|^2. \eqno(\ref{blocks}.2)$$
Now we begin to
solve the minimization problems
$$\min\limits _{{ \left \{ \begin{array}{ll}
X_{11}+X_{22}+\cdots+X_{n_1,n_1}=\rho_2 \\
\tr(X_{ii})=a_{ii},1 \leq i\leq n_1
\end{array}\right.}}
\sum\limits_{1\leq i\leq n_1}\|P_{ii}-X_{ii}\|^2\eqno(\ref{blocks}.3)$$
and $$\min\limits _{\tr(X_{ij})=a_{ij}, 1\leq i\neq j\leq n_1}
\sum\limits_{1\leq i\neq j\leq n_1}\|P_{ij}-X_{ij}\|^2, \eqno(\ref{blocks}.4)$$ respectively.

One can verify that the minimization problem (\ref{blocks}.3) is equivalent to
$$\min\limits _{\tr(X_{ii})=a_{ii}, i=2,\cdots,n_1}\|P_{11}-(\rho_2-X_{22}-\cdots-X_{n_1,n_1})\|^2
+\sum\limits_{2\leq i\leq n_1}\|P_{ii}-X_{ii}\|^{2}. \eqno(\ref{blocks}.5)$$
In fact, the equality $a_{11}=\tr(X_{11})=\tr(\rho_2-X_{22}-\cdots-X_{n_1,n_1})$ always holds if
$\tr(X_{ii})=a_{ii}, i=2,3,\cdots,n_1$
because $\rho_1$ and $\rho_2$ are density matrices, i.e., $\tr(A)=\tr(B)=1.$
Now we begin to solve (\ref{blocks}.5) instead of (\ref{blocks}.3).
Since the objective function of (\ref{blocks}.5) is a convex function and
its feasible set is a convex set, then the KKT point is the
solution of (\ref{blocks}.5). Set the Lagrangian function of (\ref{blocks}.5) is
$$L(X, \zeta)=\|P_{11}-(\rho_2-X_{22}-\cdots-X_{n_1,n_1})\|^2+
\sum\limits_{2\leq i\leq n_1}\|P_{ii}-X_{ii}\|^{2}-
\sum\limits_{2\leq i\leq n_1} \zeta_{i}(\tr(X_{ii})-a_{ii}),$$
where $\zeta=(\zeta_{2}, \zeta_{3},\cdots,\zeta_{n_1})\in \mathbb{C}^{n_1-1}.$
Hence we can derive that the optimality conditions of (\ref{blocks}.5) are $$
\left\{\begin{array}{lll} \nabla_{X_{22}}L(X, \zeta)&=& (P_{11}-\rho_2+X_{22}+\cdots+X_{n_1,n_1})
+(X_{22}-P_{22})+\frac{1}{2}\zeta_{2} I_{n_2}=0, \\
\nabla_{X_{33}}L(X, \zeta) &=& (P_{11}-\rho_2+X_{22}+\cdots+X_{n_1,n_1})+(X_{33}-P_{33})+\frac{1}{2}
\zeta_{3} I_{n_2}=0,\\
&\vdots&\\
\nabla_{X_{n_1,n_1}}L(X, \zeta) &=&(P_{11}-\rho_2+X_{22}+\cdots+X_{n_1,n_1})+(X_{n_1,n_1}-P_{n_1,n_1})+\frac{1}{2}
\zeta_{n_1} I_{n_2}=0\\
\nabla_{\zeta_{i}}L(X, \zeta) &=& \tr(X_{ii})-a_{ii}=0,\ \ 2\leq i\leq n_1,
\end{array}\right. $$
which imply that the KKT points of (\ref{blocks}.5) are
$$X_{ii}=T+P_{ii}+\frac{1}{n_2}\left(a_{ii}-\tr(P_{ii})\right)I_{n_2},\
\ 1\leq i\leq n_1. \eqno(\ref{blocks}.6)$$
where
\[T  =  \frac{1}{n_1}\left(\rho_2-\sum\limits_{i=1}^{n_1}P_{ii}\right)-\frac{1}{n_1n_2}\tr
\left(\rho_2-\sum\limits_{i=1}^{n_1}P_{ii}\right)I_{n_2}
 =  \frac{1}{n_1}\left(\rho_2-\sum\limits_{i=1}^{n_1}P_{ii}\right) +\frac{1-\tr(P)}{n_1n_2}I_{n_2}
\]
These KKT points are also the unique solution of the minimization problem (\ref{blocks}.5).
Noting that (\ref{blocks}.3) and (\ref{blocks}.5) are equivalent, then (\ref{blocks}.6) are also the unique solution of (\ref{blocks}.3).

Next we will solve the minimization problem (\ref{blocks}.4). Since the objective function of (\ref{blocks}.4)
is a convex function and its feasible set is a convex set, then the KKT point is the solution
of problem (\ref{blocks}.4). Set the Lagrangian function of (\ref{blocks}.4) is $$L(X, \mu)=\sum\limits_{1\leq i\neq j\leq n_2}\|P_{ij}
-X_{ij}\|^{2}- \sum\limits_{1\leq i\neq j\leq n_2} \mu_{ij}(\tr(X_{ij})-a_{ij}),$$
where $\mu=(\mu_{12}, \mu_{13}, \cdots, \mu_{(n_1,n_1-1)})\in \mathbb{C}^{\frac{1}{2}n_1(n_1-1)}.$

By the optimality conditions $$
 \left \{ \begin{array}{rll} \nabla_{X_{ij}}L(X, \mu)
 &=& -2P_{ij}+2X_{ij}-\mu_{ij} I_{n_2}=0,\ \ 1\leq i\neq j\leq n_1, \\
\nabla_{\mu_{ij}}L(X, \mu) &=& \tr(X_{ij})-a_{ij}=0,\ \ 1\leq i\neq j\leq n_1,
\end{array}\right.$$ we obtain the KKT point of the minimization problem (6.4) are
$$ X_{ij} = P_{ij}+\frac{1}{n_2}(a_{ij}-\tr(P_{ij}))I_{n_2},
\ \ 1\leq i\neq j\leq n,\eqno(\ref{blocks}.7)$$
which are also the unique solution of (\ref{blocks}.4).

Combining (\ref{blocks}.6), (\ref{blocks}.7) and (\ref{blocks}.2)
we see that the projection operator of $P$ onto the set $\Omega_{1}$ is indeed given by equation (\ref{blocks}).
\qed\bigskip

%\noindent\textbf{Proof of Lemma \ref{rank1k}:} Let $r=rank(X)$ and let $U$ be a unitary matrix such that $U^*XU=\diag(x_1,\ldots, x_r,0,\ldots,0)$. Let $\omega_k$ be a principal $k^{th}$ root of unity.
%\[\tilde{u}_i=[\omega_k^{(i-1)(j-1)}\sqrt{x_j}]_{j=1}^{n} \quad \mbox{ and } u_i=U\tilde{u}_i \]
%The conclusion follows. \qed \bigskip

%\begin{lemma} \label{rank1k}
%For any $X\in D_n$ and $k\geq \rank(X)$, there exists unit vectors $u_1,\ldots, u_k\in \mathcal{C}^n$ such that \[X=\frac{1}{k}\sum\limits_{i=1}^{k} u_iu_i^*\]
%\end{lemma}

\section{Proof of Proposition \ref{rankk}}\label{proofs2}

As discussed in section \ref{prelim}, we can assume without loss of generality that $n_1\leq n_2$ and that $\rho_1=(a_1,\ldots, a_{n_1})$ and $\rho_2=\diag(b_1,\ldots, b_{n_2})$ are positive definite. Let $n_2\leq k\leq n_1+n_2-1$. 
One can verify that $\rho_1=\frac{1}{k}\sum\limits_{i=1}^{k} x_ix_i^*$ and $\rho_2=\frac{1}{k}\sum\limits_{i=1}^{k} y_iy_i^*$, where $\omega_k$ is a principal $k^{th}$ root of unity and
\[x_i=[\omega_k^{(j-1)(i-1)}\sqrt{a_j}]_{j\in \underline{\mathbf{n_1}}} \quad \mbox{ and }\quad y_i=[\omega_k^{(l-1)(i-1)}\sqrt{b_l}]_{l\in \underline{\mathbf{n_2}}} \]
are unit vectors for $i=1,\ldots, k$. Define $\rho=\sum\limits_{i=1}^k \frac{1}{k} (x_ix_i^*\otimes y_iy_i^*)$. It follows that $\tr_1(\rho)=\rho_2$ and $\tr_2(\rho)=\rho_1$. 

To show that $\mbox{rank}(\rho)=k$, note that $\rho=\frac{1}{k}PP^*$, where $P$ be the $n_1n_2\times k$ matrix \[P =\begin{bmatrix}
x_1 \otimes y_1 & x_2 \otimes y_2 & \cdots & x_k\otimes y_k
\end{bmatrix}.\]
Let \[F=\begin{bmatrix}
1 & 1 &  \cdots & 1 \\
1 & \omega_k &\cdots  & \omega_k^{k-1}\\
& \vdots  & \ddots & \vdots \\
1 & \omega_k^{n-1} & \cdots & \omega_k^{(k-1)(n_2-1)}
\end{bmatrix}\quad \mbox{ and } \quad D=\diag(1,\omega_k, \omega_k^2,\ldots, \omega_k^{k-1}) \]
Then \[P=\diag\left(\sqrt{a_1},\ldots, \sqrt{a_{n_1}})\otimes \diag(\sqrt{b_1}, \ldots, \sqrt{b_{n_2}}\right)\begin{bmatrix}
F\\
FD\\
\vdots \\
FD^{k-1}
\end{bmatrix} \]
Note that $FD^i$ consists of the $(1+i)^{th}$ up to the $(n_2+i)^{th}$ row of the discrete $k\times k$ Fourier matrix, which is a unitary matrix. Hence, $P$ has $k$  linearly independent rows consisting of rows $1,\ldots, n_2, 2n_2,3n_2,\ldots, (k-n_2+1)n_2$. 
Counting all the linearly independent rows of $P$, we get that $\rank(P)=\rank(\rho)=k$.\qed

\section{Proof of Theorem \ref{inductrank}}\label{proofs3}

Assume without loss of generality that $n_1\leq n_2$,  $\rank(\rho_1)=n_1$, $\rank(\rho_2)=n_2$ and that
$\rho_1 = \diag(a_1, \dots, a_{n_1}) \mbox{ and 
}\rho_2 = \diag(b_1, \dots, b_{n_2}).$

If $n_1=1$ (or $n_2=1$), then $\cS(\rho_1,\rho_2)=\{\rho_1\otimes \rho_2\}$ and $\rank(\rho_1\otimes \rho_2)=n_2$ (or $\rank(\rho_1\otimes \rho_2)=n_1$). 

We will prove the general statement using induction on $n_1+n_2$. By the preceding remark, the statement holds for $n_1+n_2=2$. Now, suppose that the statement holds (i.e. for any $n_2\leq k \leq n_1n_2$, there exists a rank $k$ solution $\rho\in\cS(\rho_1,\rho_2)$) when $n_1+n_2$ satisfies $2\leq n_1+n_2< r$.  

Consider the case $n_1+n_2=r$. By Proposition \ref{rankk}, for any $n_1\leq k\leq n_1+n_2-1$ there is a rank $k$ solution $\rho\in \mathcal{D}_{n_1n_2}$ such that $\tr_1(\rho)=\rho_2$ and $\tr_2(\rho)=\rho_1$.
\begin{enumerate}
\item If $n_1=1$, then we are done.
\item If $1<n_1<n_2$, then  using the induction hypothesis, we know that for any $ n_2-1\leq k\leq n_1(n_2-1)$ there is a rank $k$ density matrix $\rho $ such that 
\[\tr_1(\rho)=\frac{1}{1-b_{n_2}}\diag(b_1,\ldots, b_{n_2-1},0) \mbox{ and } \tr_2(\rho)=\diag(a_1,\ldots,a_{n_1}). \]
Now, let $\hat{\rho}=(1-b_{n_2})\rho+\diag(a_1,\ldots,a_{n_1})\otimes \diag(0,\ldots, 0, b_{n_2})\in \cS(\rho_1,\rho_2)$. The fact that $\rank(\hat{\rho})=\rank(\rho)+n_1$ is evident from its block structure. Thus, $\rho$ can be chosen so that $\hat{\rho}$ has rank ranging from $n_1+n_2-1$ to $n_1n_2$. Together with Proposition \ref{rankk}, this shows that there is a solution of rank $k$ for any $n_2\leq k\leq n_1n_2$. 
\item If $1<n_1=n_2$, then using the induction hypothesis, we know that for any $ n_2\leq k\leq (n_1-1)n_2  $ there is a rank $k$ density matrix $\rho $ such that 
\[\tr_1(\rho)=\diag(b_1,\ldots, b_{n_2}) \mbox{ and } \tr_2(\rho)=\frac{1}{1-a_{n_1}}\diag(a_1,\ldots,a_{n_1-1},0).\]
One can verify that $\hat{\rho}=(1-a_{n_1})\rho+\diag(b_1,\ldots, b_{n_2})\otimes \diag(0,\ldots,0,a_{n_1})\in\cS(\rho_1,\rho_2)$ and $\rank(\hat{\rho})=\rank(\rho)+n_2$. Thus, $\hat{\rho}$ can have rank ranging from $2n_2$ to $n_1n_2$. Together with Proposition \ref{rankk}, this shows that there is a solution of rank $k$ for any $n_2\leq k\leq n_1n_2$. 
\end{enumerate}
By the principle of mathematical 
induction, we see that the theorem holds for all $1 \leq n_1 \leq n_2$. \qed\bigskip

\section{Proof of Proposition \ref{interlacethm}}\label{proofs4}
 Note that in Algorithm \ref{alginterlace}, $A_k$ and $B_k$ are positive semidefinite for every iteration $k$. The recursive process terminates at iteration $k$ when $\rank(A_k)=0$. From the construction, we get 
\[\rank(A_{k+1})=\rank(A_k)-\left(\sum_{i=1}^{p}\rank(S_i)\right)-\left(\sum_{j=1}^{q}\rank(T_j)\right) +p\] 
\[\rank(B_{k+1})=\rank(B_k)-\left(\sum_{i=1}^{p}\rank(\tilde{S}_i)\right)-\left(\sum_{j=1}^{q}\rank(\tilde{T}_j)\right) +q\]
Since $\mbox{tr}(A_k)=\mbox{tr}(B_k)$ and $A_k,B_k$ are both positive semidefinite, then there exists $a_{i_1},a_{i_2},b_{j_1},b_{j_2}$ such that $a_{i_1}\geq b_{j_1}$ and $b_{j_2}\geq a_{j_2}$. Thus, $p,q\geq 1$. Hence, $\rank(A_{k+1})< \rank(A_k)$ and $\rank(B_{k+1})<\rank(B_k)$ so that the process terminates after at most $\max\{\rank(\rho_1),\rank(\rho_2)\}$ steps. Clearly $C_i$ and $\tilde{C}_i$ are positive semidefinite and are isospectral and $\rho_1=C_1+\cdots +C_k$ and $\rho_2=\tilde{C}_1+\cdots+\tilde{C}_k$. It follows from Theorem \ref{pure} that $\tr_1(w_iw_i^*)=\tilde{C}_i$ and $\tr_2(w_iw_i^*)=C_i$ for $i=1,\ldots k$. Thus, $\rho\in \mathcal{S}(\rho_1,\rho_2)$.

If $a_{i_1}\geq b_{j_1}\geq \cdots \geq a_{i_l}\geq b_{j_l}$ (or $b_{j_1}\geq a_{i_1}\geq \cdots \geq b_{j_l}\geq a_{i_l}$)  for some distinct indices $i_1,\ldots,i_{l+1}$ and distinct $j_1,\ldots,j_{l+1}$, then $\rho_1=C_1+A_1$ and $\rho_2=\tilde{C}_1+B_1$ where $\rank(A_1)\leq \rank(\rho_1)-l$ and $\rank(B_1)\leq \rank(\rho_2)-l$. 
\qed

\section{Proof of Proposition \ref{lowrankCthm}}\label{proofs5}

As discussed in section \ref{prelim}, we can assume without loss of generality that $n_1\leq n_2$ and 
\[\rho_1=\diag(a_1,\ldots,a_{n_1}) \mbox{ and } \rho_2=\diag(b_1,\ldots, b_{n_2}),\] where $a_1\geq a_2\geq\cdots \geq a_{n_1}$ and $b_1\geq b_2\geq \cdots \geq b_{n_2}$. For any $i=1,\ldots,n_1$, define $c_i=\min\{a_i,b_i\}$ and $c_{n_1+1}=\cdots=c_{n_2}=0$ and define $C_1=\diag(c_1,\ldots, c_{n_1})$ and $\tilde{C}_1=\diag(c_1,\ldots, c_{n_2})$. Then $\rho_1-C_1$ is positive semidefinite and has rank at least one less than $\rank(\rho_1)$. Similarly, $\rho_2-\tilde{C}_1$ is positive semidefinite and has rank at least one less than $\rank(\rho_2)$. We can replace $\rho_1$ and $\rho_2$ by $\rho_1-C_1$ and $\rho_2-\tilde{C}_1$ and repeat the above process until both matrices become zero. This process will take at most $k=\max\{\rank(\rho_1),\rank(\rho_2)\}$ steps because the rank of $\rho_1$ and $\rho_2$ are reduced by at least one in each step. At the end of this process, we will be able to write $\rho_1$ and $\rho_2$ as  $\rho_1=C_1+\cdots +C_k$ and $\rho_2=\tilde{C}_1+\cdots+\tilde{C}_k$ such that for each $i$, 
\[C_i=\diag(c^{(i)}_{1},\ldots, c^{(i)}_{n_1}) \mbox{ and } \tilde{C}_i=\diag(c^{(i)}_{\sigma_i(1)},\ldots, c^{(i)}_{\sigma_i(n_2)})\] for some permutation map $\sigma_i:\underline{\mathbf{n}_2}\longrightarrow \underline{\mathbf{n}_2}$. Equivalently, we have partitioned the eigenvalues of $\rho_1$ and $\rho_2$ such that 
\[a_i=\sum\limits_{r=1}^k c^{(r)}_{i} \ \forall i\in\underline{\mathbf{n_1}} \mbox{ and } b_j=\sum\limits_{r=1}^k c^{(r)}_{\sigma_r(j)} \ \forall j\in\underline{\mathbf{n_2}}\] 
Observe that in this scheme, it is true that if $c^{(t)}_{i}\neq 0$, either $c^{(r)}_{i}=0$ for all $r\geq t$ or $c^{(r)}_{\sigma_{r}(\sigma_t^{-1}(i))}=0$ for all $s\geq t$. That is, $c^{(t)}_{i}$ is either the last nonzero summand of $a_i$ or the last nonzero summand for $b_{\sigma_t^{-1}(i)}$. 

Let $\rho=w_1w_1^*+\cdots + w_kw_k^*$, where $w_r=\sum_{j=1}^{n_1} \sqrt{c^{(r)}_{j}} (e_j^{(n_1)}\otimes e_{\sigma^{-1}(j)}^{(n_2)})$.
%From this construction, we can also deduce that if $c_{i_j}, c_{k_j}\neq 0$, then $\sigma_i(j)\neq \sigma_k(j)$. 
Note that for $p\neq q$,
\[\begin{array}{lcl}
 w_p^*w_q & =  & \sum\limits_{j,\ell=1}^{n_1} \sqrt{c^{(p)}_{j}c^{(q)}_{\ell}}\left(e^{(n_1)*}_je^{(n_1)}_{\ell}\otimes  e^{(n_2)*}_{\sigma_p^{-1}(j)}e^{(n_2)}_{\sigma_q^{-1}(\ell)}\right)  \medskip\\
 & = & \sum\limits_{j=1}^{n_1} \sqrt{c^{(p)}_{j}c^{(q)}_{j}}\left(e_{\sigma_p^{-1}(j)}^{(n_2)*}e_{\sigma_q^{-1}(j)}^{(n_2)}\right) \quad  = \quad  \sum\limits_{\stackrel{j=1}{\sigma_p^{-1}(j)=\sigma_q^{-1}(j)}}^{n_1} \sqrt{c^{(p)}_{j}c^{(q)}_{j}}
\end{array}
\]
Suppose $q>p$ and $\sigma^{-1}_p(j)=\sigma^{-1}_q(j)$. By construction, if $c^{(p)}_{j}\neq 0$, then $c^{(q)}_{j}=c^{(q)}_{\sigma_q\sigma_p^{-1}(j)}=0$. Similarly, if $p>q$ and $\sigma^{-1}_p(j)=\sigma^{-1}_q(j)$, then either $c^{(q)}_j=0$ or $c^{(p)}_{j}=c^{(p)}_{\sigma_p\sigma_q^{-1}(j)}=0$. Thus $w_1,\ldots, w_k$ form an orthogonal basis. This means that for $r=1,\ldots k$, $\lambda_r=||w_r||^2=c^{(r)}_{1}+\cdots + c^{(r)}_{n_1}$, (together with $n_1n_2-k$ more zeros) are the eigevalues of $\rho$. 

%From \cite{Kly1}, we have that $\mbox{tr}(C_1)=\sum\limits_{i=1}^{n_1}\min\{a_i,b_i\}$.  

Now, suppose $\sigma\in \mathcal{S}(\rho_1,\rho_2)$ with spectral decomposition $s_1x_1x_1^*+\cdots + s_{N}x_{N}x_{N}^*$. Then  
\[\rho_1=s_1\tr_2(x_1x_1^*)+\cdots + s_N\tr_2(x_Nx_N^*)\quad  \mbox{ and }\quad \rho_2=s_1\tr_1(x_1x_1^*)+\cdots + s_N\tr_1(x_Nx_N^*)\]
Hence $\rho_1-s_1\tr_2(x_1x_1^*) $ and $\rho_2-s_1\tr_1(x_1x_1^*)$ are positive semidefinite. Let $c_1\geq \cdots\geq  c_k$ be the nonzero eigenvalues of $s_1\tr_2(x_1x_1^*)$, which are also 
the nonzero eigenvalues of $s_1\tr_1(x_1x_1^*)$. Then using Lidskii's inequalities, we get $c_i\leq \min\{a_i,b_i\}$ for $i=1,\ldots, k$. Thus,
\[|| \sigma||_{2}=s_1=\sum\limits_{i=1}^k c_i\leq \sum\limits_{i=1}^k\min\{a_i,b_i\} \leq \sum\limits_{i=1}^{\min\{n_1,n_2\}}\min\{a_i,b_i\}=|| \rho||_{2}\]
\qed\bigskip

\section{Proof of Proposition \ref{multiprob} and \ref{multiprojform}}\label{proofs6}

Note that the condition $\tr_{J_i^c}(\rho)=\rho_{J_{i}}$ can be written as a set of
linear constraints of the form $A_j x=b_j$ by vectorizing $\rho$ into $x\in \mathbb{R}^{n}$ and
$\rho_{J_{i}}$ into $b_i\in \mathbb{R}^m$. First, we look at the projection of a given
$\hat{x}\in \mathbb{R}^n$ onto the set of solutions of a linear constraint of the form $Ax=b$,
where $A$ is an $m\times n$ real matrix. For this, we need the Moore-Penrose inverse of $A$,
denoted by $A^{+}$, which is the unique $n\times m$ matrix satisfying the following four
conditions:\smallskip\\
\centerline{
(a) $AA^{+}A=A$, \quad (b)
$A^{+}AA^{+}=A^{+}$, \quad (c) $AA^{+}$ is symmetric,   \quad
(d) $A^{+}A$ is symmetric. }\medskip\\ 
It is known that \[\tilde x  = x - A^+(Ax-b) \mbox{ satisfies }
\|x-\tilde x\| \le \|x-z\| \ \hbox{ for all } z \in L = \{x \in \IR^n: Ax = b\}\neq \emptyset.\] Applying this to a linear operator $T: H_N \rightarrow H_n$ and the set $\cL = \{ \rho\in H_N: T(\rho) = \sigma\}$, we get 
\[\tilde \rho = \rho - T^+(T(\rho) - \sigma) \mbox{ satisfies } ||\rho-\tilde{\rho}|| \leq ||\rho- X|| \mbox{ for all } X\in \mathcal{L}.\] 
Here $T^{+}$ is the unique map $T^{+}:H_n\rightarrow H_N$ satisfying the conditions:
\begin{enumerate}
\item[(a)] $TT^{+}T=T$,
\item[(b)] $T^{+}TT^{+}=T^{+}$,
\item[(c)] $\mbox{tr}(TT^{+}(X)^*Y) =\mbox{tr}(X^* (TT^{+}(Y)))$ for all $X,Y\in H_n$,
\item[(d)] $\mbox{tr}(T^{+}T(X)^*Y) =\mbox{tr}(X^* (T^{+}T(Y)))$ for all $X,Y\in H_N$.
\end{enumerate}

Let $T:H_{mn}\longrightarrow H_{n}$ such that $T(\rho)=\tr_1(\rho)$ and
$S:H_{n}\longrightarrow H_{mn}$ such that $S(\sigma)=\frac{I_m}{m}\otimes B$ for all $\sigma\in H_n$. It is clear that $TST(\rho)=T(\rho)$ for all $\rho\in H_{mn}$ and $STS(\sigma)=S(\sigma)$ for any
$\sigma\in H_n$. Note that, $TS$ is the identity map, and hence a hermitian operator on $H_n$. Finally,
we show that $ST$ is a hermitian operator as follows: let $\rho,\nu\in H_{mn}$ with block structure
$\rho=[\rho_{ij}]$ and $\nu=[\nu_{ij}]$, where $\rho_{ij},\nu_{ij}\in M_{n}$.
\[\langle ST(\rho),\nu\rangle=\tr\left(\left(\frac{I_m}{m}\otimes
\tr_1(\rho)\right)\nu\right)=\tr\left(\left[\frac{\tr_1(\rho)
\nu_{ij}}{m}\right]_{ij}\right)=\tr\left(\frac{\tr_1(\rho)\tr_1(\nu)}{m}\right)\]
Similarly,
\[\langle\rho,ST(\nu)\rangle=\tr\left(\rho\left(\frac{I_m}{m}\otimes \tr_1(\nu)\right)\right)=\tr\left(\left[\frac{\rho_{ij}\tr_1(\nu)}{m}\right]_{ij}\right)=\tr\left(\frac{\tr_1(\rho)\tr_1(\nu)}{m}\right)\]
Thus $S=T^{+}$.

Now, to prove Proposition \ref{multiprob}, let $J\subseteq \underline{\mathbf{k}}$, $n_{J}=\prod\limits_{i\in J} n_i$ and $n_{J^c}=\prod\limits_{i\in J^c} n_i$ and $P_J$ be as defined in equation (\ref{permute}). Then for the partial trace operator $T_{J}: H_{n_1\cdots n_k}\longrightarrow H_{n_J}\mbox{ such that } T_J(\rho)=\tr_{J^{c}}(\rho)=\rho_{J},$  we have,  \[T_J^{+}(\sigma)=P\left(\frac{I_{n_{J^c}}}{n_{J^c}}\otimes \sigma\right)P^{T}\]
for all $\sigma\in H_{n_J}$. Therefore, the least square approximation of $Z\in H_{n_1\cdots n_k}$ in $\mathcal{L}=\{\rho\in H_{n_1\cdots n_k}:
\tr_{J^c}(\rho)=\sigma\}$ is given by
\[
\Phi_{J}(Z)=Z-T^{+}_{J}(T_J(Z)-\sigma),
= Z-P_J^{T}\left(\frac{I_{n_{J^c}}}{n_{J^c}}\otimes
(\tr_{J^c}(Z)-\sigma)\right)P_J\]
\qed

Given $Z\in M_{n_1\cdots n_k}$, denote the column vector obtained by stacking the columns of $Z$ by $\mbox{vec}(Z)$.  Then there are matrices $A_1,\ldots, A_m$ such that 
\[\mathcal{L}=\{\rho \ | A_i \mbox{vec}(\rho)=\mbox{vec}(\rho_{J_i})  \mbox{ for } i=1,\ldots, m\} \]

Proposition \ref{multiprojform} will follow directly from Proposition \ref{multiprob} and the following theorem.

\begin{theorem}\label{Amulti}
Let $A_i\in M_{n_i,N}$ and $b_i\in M_{n_i}$ for $i=1,\ldots, m$. For any $\{i_1,\ldots,i_r\} \subseteq\{1,\ldots,m\}$, denote by $A_{[i_1,\ldots, i_r]}$ the matrix whose row space is $\bigcap\limits_{j=1}^r \mbox{\rm Row}(A_{i_j})$. 
The set \[L=\{x \ |\ A_ix=b_i \mbox{ for } i =1,\ldots,m\}\] is nonempty if and only if for any subset $\{i_1,\ldots,i_r\}$ of $\{1,\ldots,m\}$, the projection of $b_{i_s}$ onto $\bigcap\limits_{j=1}^r \mbox{\rm Row}(A_{i_j})$ is constant for all $s=1,\ldots r$. In this case, denote this projection by $b_{[i_1,\ldots,i_r]}$. 
Then the least square projection of $z\in \mathbb{C}^N$ onto $L$ is given by
\[\tilde{z}=z+\sum\limits_{r=1}^m (-1)^r \sum\limits_{\{i_1,\ldots,i_r\}\subseteq \{1,\ldots,m\}}A_{[i_1,\ldots,i_r]}^{+}\Big(A_{[i_1,\ldots,i_r]}x-b_{[i_1,\ldots,i_r]}\Big) \]
\end{theorem}

\noindent\textbf{Proof:} We will prove this theorem by induction. 

First, we consider the case when $m=2$. Let $V=\begin{pmatrix}
V_1^{T} &
V_2^{T} &
V_3^{T}
\end{pmatrix}^{T}$ such that the rows of $V_1$ form an orthonormal basis for $\Row(A_1)\cap \Row(A_2)^{\perp}$, the rows of $V_2$ form an orthonormal basis for $\Row(A_1)\cap \Row(A_2)$ and the rows of $V_3$ form an orthonormal basis for $\Row(A_2)\cap \Row(A_1)^{\perp}$. Then for some unitary $U_1=\begin{pmatrix}
U_{11}\\
U_{21}\end{pmatrix}\in M_{n_1}$ and $U_2=\begin{pmatrix}
U_{12}\\
U_{22}\end{pmatrix}\in M_{n_2}$, we have
\[\begin{pmatrix}
A_1\\
A_2
\end{pmatrix} = (U_1^*\oplus U_2^*) \begin{pmatrix}
C_1 & 0 & 0\\
0 & C_2 & 0\\
0 & C_2 & 0\\
0 & 0 & C_3
\end{pmatrix}V \]
Thus,
\[\begin{array}{lcl}
\begin{pmatrix}
A_1\\
A_2
\end{pmatrix}^{+} & = &  V^*\begin{pmatrix}
C_1^{+} & 0 & 0 & 0\\
0 & \nicefrac{C_2^{+}}{2} &  \nicefrac{C_2^{+}}{2} & 0\\
0 & 0 & 0 & C_3^{+1}
\end{pmatrix}(U_1 \oplus U_2)\bigskip\\
& = & \begin{pmatrix}
A_1^{+} &
A_2^{+}
\end{pmatrix}-\frac{1}{2}\begin{pmatrix}
A_1^{+}P_1^* &
A_2^{+}P_2^* &
\end{pmatrix}
\end{array}\]
where $P_1=U_1^* \begin{pmatrix}
0 & 0 \\
0 & I
\end{pmatrix}U_1$ is the projection from $\Row(A_1)$ to $\Row(A_1)\cap \Row(A_2)$ and $P_2=U_2^* \begin{pmatrix}
I & 0 \\
0 & 0
\end{pmatrix}U_2$ is the projection from $\Row(A_2)$ to $\Row(A_1)\cap \Row(A_2)$.
 Note that \[A_1^{+}P_1^*A_1=V\begin{pmatrix}
0 & 0\\
 0& C_2^{+}\\
 0 & 0
 \end{pmatrix}\begin{pmatrix}
C_1& 0 & 0\\
 0& C_2 & 0\\
 \end{pmatrix}V^*=V\begin{pmatrix}
0 & 0\\
 C_2^{+}& 0\\
 0 & 0
 \end{pmatrix}\begin{pmatrix}
0 & C_2 & 0\\
 0& 0 & C_3\\
 \end{pmatrix}V^*=A_2^{+}P_2^*A_2:=A_{[1,2]}.\]
 If $L\neq \emptyset$, then there must be $\tilde{x}$ such that $A_1\tilde{x}=b_1$ and
 $A_2\tilde{x}=b_2$. Thus
 $A_1^{+}P_1^*b_1=A_2^{+}P_1^*A_1\tilde{x}=A_2^{+}P_2^*A_2\tilde{x}=A_2^{+}P_2^*b_2:b_{[1,2]}$.
Hence, the least square approximation of a given $x\in\mathbb{R}^n$ on the set $L$ is given by
\[\begin{array}{lcl}
\tilde{x} & = & x-\begin{pmatrix}
A_1\\
A_2
\end{pmatrix}^{+}\left(\begin{pmatrix}
A_1\\
A_2
\end{pmatrix}x-\begin{pmatrix}
b_1\\
b_2
\end{pmatrix}\right)\bigskip\\
& = & x-A_1^{+}(A_1x-b_1)-A_2^{+}(A_2x-b_2)+\frac{1}{2}A_1^{+}P_{1}^*(A_1x-b_1)+
\frac{1}{2}A_2^{+}P_{2}^*(A_2x-b_2)\medskip\\
& = & x-A_1^{+}(A_1x-b_1)-A_2^{+}(A_2x-b_2)+A_{[1,2]}^{+}(A_{[1,2]}x-b_{[1,2]})
\end{array}
 \]
This proves the theorem for the case $m=2$. 

Now, suppose it is true for $m=2,\ldots, s-1$. The least square approximation of a given $x\in \mathbb{R}^N$ on $\mathcal{L}$
is given by
$$\hat{x}=x-\begin{pmatrix}
A_1\\
A_2\\
\vdots\\
A_{s}
\end{pmatrix}^{+}\left(\begin{pmatrix}
A_1\\
A_2\\
\vdots\\
A_{s}
\end{pmatrix}x-\begin{pmatrix}
b_1\\
b_2\\
\vdots\\
b_{s}
\end{pmatrix} \right).$$
From the $m=2$ case, we have
$$\hat{x}=x-\begin{pmatrix}
A_1\\
\vdots\\
A_{s-1}
\end{pmatrix}^{+}\left(\begin{pmatrix}
A_1\\
\vdots\\
A_{s-1}
\end{pmatrix}x-\begin{pmatrix}
b_1\\
\vdots\\
b_{s-1}
\end{pmatrix} \right)-A_s^{+}(A_sx-b_s)+\begin{pmatrix}
A_{[1,s]}\\
\vdots\\
A_{[s-1,s]}
\end{pmatrix}^{+}\left(\begin{pmatrix}
A_{[1,s]}\\
\vdots\\
A_{[s-1,s]}
\end{pmatrix}x-\begin{pmatrix}
b_{[1,s]}\\
\vdots\\
b_{[s-1,s]}
\end{pmatrix}\right),$$
Apply the induction hypothesis to get
\[\begin{array}{lcl}
y_1 & = & x-\begin{pmatrix}
A_1\\
\vdots\\
A_{s-1}
\end{pmatrix}^{+}\left(\begin{pmatrix}
A_1\\
\vdots\\
A_{s-1}
\end{pmatrix}x-\begin{pmatrix}
b_1\\
\vdots\\
b_{s-1}
\end{pmatrix} \right)\\
& = &  x+\sum\limits_{r=1}^{s-1} (-1)^r \sum\limits_{\{i_1,\ldots,i_r\}\subseteq \{1,\ldots,s-1\}}A_{[i_1,\ldots,i_r]}^{+}\Big(A_{[i_1,\ldots,i_r]}x-b_{[i_1,\ldots,i_r]}\Big)
\end{array}\]
\[\begin{array}{lcl}
y_2 & = & x-\begin{pmatrix}
A_{[1,s]}\\
\vdots\\
A_{[s-1,s]}
\end{pmatrix}^{+}\left(\begin{pmatrix}
A_{[1,s]}\\
\vdots\\
A_{[s-1,s]}
\end{pmatrix}x-\begin{pmatrix}
b_{[1,s]}\\
\vdots\\
b_{[s-1,s]}
\end{pmatrix}\right)\medskip\\
& = &  x+\sum\limits_{r=1}^{s-1} (-1)^r \sum\limits_{\{i_1,\ldots,i_r\}\subseteq \{1,\ldots,s-1\}}A_{[i_1,\ldots,i_r,s]}^{+}\Big(A_{[i_1,\ldots,i_r,s]}x-b_{[i_1,\ldots,i_r,s]}\Big)
\end{array}\]
Then $\hat{x}=y_1-y_2+x-A_s^{+}(A_sx-b_s)$, which gives the desired equation.\qed

\end{document}